\documentclass{aa}  

\usepackage{graphicx}
\usepackage{txfonts}
\begin{document} 
%
\newcommand{\herschel}{\emph{Herschel}}
\newcommand{\spitzer}{\emph{Spitzer}}
\newcommand{\planck}{\emph{Planck}}
\newcommand{\akari}{\emph{Akari}}
\newcommand{\iras}{IRAS}
\newcommand{\iso}{ISO}
\newcommand{\alma}{ALMA}
\newcommand{\galex}{GALEX}
\newcommand{\wise}{WISE}
\newcommand{\lsun}{\mbox{L$_\odot$}}
\newcommand{\msun}{\mbox{M$_\odot$}}
\newcommand{\msunyr}{\mbox{M$_\odot$~yr$^{-1}$}}
\newcommand{\rsun}{\mbox{R$_\odot$}}
\newcommand{\lbol}{$L_{\rm bol}$} 
\newcommand{\lfir}{$L_{\rm FIR}$} 
\newcommand{\lir}{$L_{\rm IR}$} 
\newcommand{\reblue}{$R_{\rm e,70}$} 
\newcommand{\regreen}{$R_{\rm e,100}$} 
\newcommand{\rered}{$R_{\rm e,160}$} 
\newcommand{\sigfir}{$\Sigma_{\rm FIR}$} 
\newcommand{\sigsfr}{$\Sigma_{\rm SFR}$} 
\newcommand{\tdust}{$T_{\rm dust}$}
\newcommand{\mstellar}{$M_{\ast}$}     
\newcommand{\ergs}{erg~s$^{-1}$}    
\newcommand{\ergscm}{erg~s$^{-1}$~cm$^{-2}$} 
\newcommand{\mum}{\mbox{$\mu$m}}
\newcommand{\cii}{[C{\sc ii}]}

\makeatletter
\newcommand{\pushright}[1]{\ifmeasuring@#1\else\omit\hfill$\displaystyle#1$\fi\ignorespaces}
\newcommand{\pushleft}[1]{\ifmeasuring@#1\else\omit$\displaystyle#1$\hfill\fi\ignorespaces}
\makeatother

   \title{The far-infrared emitting region in local 
galaxies and QSOs: Size and scaling relations}


   \author{D.~Lutz \inst{1}
          \and S.~Berta\inst{1}
          \and A.~Contursi\inst{1}
          \and N.M.~F\"orster Schreiber\inst{1}
          \and R.~Genzel\inst{1}
          \and J.~Graci\'a-Carpio\inst{1}
          \and R.~Herrera-Camus\inst{1}
          \and H.~Netzer\inst{2}
          \and E.~Sturm\inst{1}
          \and L.J.~Tacconi\inst{1}
          \and K.~Tadaki\inst{1}
          \and S.~Veilleux\inst{3}}

   \institute{Max-Planck-Institut f\"ur extraterrestrische Physik,
              Giessenbachstra\ss\/e, 85748 Garching, Germany\\
              \email{lutz@mpe.mpg.de}
         \and
             School of Physics and Astronomy, Tel Aviv University, 
             Tel Aviv 69978, Israel
         \and
             Department of Astronomy, University of Maryland, 
             College Park, MD 20742, USA
             }

   \date{Received 6 November 2015 ; accepted 13 May 2016}

  \abstract
   {We use \herschel\ 70 to 160~\mum\ images to study the size of the 
far-infrared emitting region in about 400 local galaxies and quasar (QSO) 
hosts. The 
sample includes normal `main-sequence' star-forming galaxies, as well as 
infrared luminous galaxies and Palomar-Green QSOs, with different levels and 
structures of star formation.  
Assuming Gaussian spatial distribution of the far-infrared (FIR) emission, 
the excellent stability of the \herschel\ point spread function (PSF) enables 
us to measure sizes well below the PSF width, by
subtracting widths in quadrature. We derive scalings of FIR size 
and surface brightness of local galaxies with FIR luminosity, with distance 
from the star-forming main-sequence, and with FIR color. 
Luminosities \lfir\/$\sim$10$^{11}$~\lsun\ can be 
reached with a variety of structures spanning 2 dex in size. 
Ultraluminous \lfir\/$\gtrsim$10$^{12}$~\lsun\  
galaxies far above the main-sequence inevitably have small 
\reblue\/$\sim$0.5~kpc FIR emitting regions with large 
surface brightness, and can be close to optically thick in the FIR on average 
over these regions. 
Compared to these local relations, 
first \alma\ sizes for the dust emission regions in high redshift galaxies, 
measured at somewhat longer rest
wavelengths, suggest larger sizes at the same IR luminosity.
We report a remarkably tight relation with 0.15~dex scatter between FIR 
surface 
brightness and the ratio of \cii~158~\mum\ emission and FIR emission -- the 
so-called \cii-deficit is more tightly linked to surface brightness than
to FIR luminosity or FIR color. 
Among 33 z$\leq$0.1 PG QSOs with typical \lfir\//$L_{\rm Bol,AGN}\approx$0.1, 
19 have a measured 70~\mum\ half light radius, 
with median \reblue\/=1.1~kpc. 
This is consistent with the
FIR size for galaxies with similar \lfir\ but lacking a QSO, in accordance 
with a scenario where the rest FIR emission of these types of QSOs is, in most 
cases, due to host star formation.}


   \maketitle
%

\section{Introduction}

The rate of star formation (SFR) is a primary characterizing quantity of any 
galaxy,
and its spatial distribution holds important clues about its driving 
mechanisms. Over recent years, it has been established that most stars form in
galaxies on a star-forming main-sequence (MS) in the M$_*$-SFR plane 
(see, e.g., \citet{brinchmann04} and \citet{schiminovich07} for the local 
universe and many works for redshifts up 
to 2 and more). These main-sequence galaxies are typically disk dominated 
objects \citep[e.g.,][]{wuyts11b,wisnioski15},
and existence and tightness of the star-forming main-sequence imply a rather 
steady star formation history in these objects. On the other hand, the most 
intense star formation in the local universe occurs in (ultra)luminous 
infrared galaxies ((U)LIRGs) that are closely related to galaxy interaction 
or mergers 
\citep[][and references therein]{sanders96}. These objects are well 
above the main-sequence, and their extreme star formation is certainly
episodic. Understanding the role of steady processes, dramatic mergers, and 
the impact of active galactic nuclei on the level and distribution of star 
formation, and on 
its ultimate quenching, is a centerpiece of galaxy evolution studies.

Star formation in most local galaxy disks can be traced well by a variety of 
UV to near-IR indicators. This gets difficult if not impossible in the heavily
dust-obscured IR luminous systems, where the short wavelength tracers 
reach only a surface layer. Longer wavelength tracers have been used
instead to map star formation at sufficient spatial resolution. Resolution 
should be a few arcsec, or better reach down to sub-arcsec levels for 
optimal study of local 
IR-luminous systems. These tracers include mostly radio continuum 
\citep[e.g.,][]{condon90}, mid-infrared continuum and emission features 
\citep[e.g.,][]{soifer00,soifer01,diazsantos10,diazsantos11}, but also 
mm interferometry to the
extent one can constrain the distribution of star formation from the 
distribution of dust or gas mass \citep[e.g.,][]{sakamoto99}. Each of these 
approaches has its particular 
strengths and weaknesses, related to the tightness of the link between the 
particular tracer and SFR, to the possible contamination by other sources 
such as AGN or dust heated by old stars, as well as to technicalities of 
available beam 
sizes, coverage of all relevant spatial frequencies in interferometry, 
and size of samples that can be realistically
observed. But a clear finding is that some IR-luminous systems are 
characterized by very compact star formation \citep[e.g.,][]{condon91}.  

A powerful approach uses high spatial resolution measurements of the 
far-infrared
emission near the $\sim$100~\mum\ SED peak where the energy of obscured
star formation is re-radiated by the dust `calorimeter'. Because of the
favourable contrast of this peak to the emission that may be heated by an 
AGN proper, 
this also is arguably the best chance to map star formation in the 
presence of fairly powerful AGN. But sensitive 
far-infrared interferometry will not be available soon, and the first 
cryogenic space telescopes such as \iras\/, \iso\/, \spitzer\/,
and \akari\ lacked
the necessary spatial resolution in the far-infrared. The best tool for the
foreseeable future is the \herschel\/\footnote{Herschel is an 
  ESA space observatory with science instruments provided by European-led 
  Principal Investigator consortia and with important participation from 
  NASA.} \citep{pilbratt10} database. While
the full width at half maximum (FWHM) of the point spread function (PSF) at 
the shorter of the 70, 100 and 160~\mum\ wavelengths
covered by \herschel\/-PACS \citep{poglitsch10} is of order 5\arcsec\/, 
characteristic
source sizes below that scale are still derivable from their broadening 
effect, given the intrinsic stability of the PSF. The present paper attempts
this for a sample of $\sim$400 galaxies, including normal near 
main-sequence as well as IR-luminous galaxies, and QSO hosts. This approach 
uses a tracer that is closely linked to SFR and can be applied homogeneously 
to hundreds of sources, but only delivers a characteristic size rather than
detailed structure that would typically need sub-arcsecond interferometry.
We adopt an $\Omega_m =0.3$, $\Omega_\Lambda =0.7$, and 
$H_0=70$ km\,s$^{-1}$\,Mpc$^{-1}$ cosmology, redshift-independent distances
for some very local galaxies, a \citet{chabrier03} IMF, a conversion
${\rm SFR} = 1.9\times 10^{-10}L_{FIR=40-120\mum}$ as appropriate for the
\citet{kennicutt98} conversion corrected to Chabrier IMF, and a ratio 1.9
of 8--1000~\mum\ and 40--120~\mum\ luminosity that is typical for the non-QSO 
galaxies in our sample.


\section{Data and analysis methods}

We use 70, 100, and 160~\mum\ images from scan maps obtained with
PACS \citep{poglitsch10} on board \herschel\ \citep{pilbratt10}, 
collecting archival data from various projects. In order to cover a wide
range of galaxy properties, we first obtain an IR-selected local sample 
ranging from
normal galaxies up to (ultra)luminous infrared galaxies. For
that purpose, we 
searched the \herschel\ archive for all cz$\geq$2000~km/s objects from the
IRAS Revised Bright Galaxy Sample \citep[RBGS,][]{sanders03}. Various proposals
included some RBGS galaxies, overall covering a wide range of IR luminosities from
normal to ULIRG. Coverage of the RBGS sample by \herschel\ is excellent at LIRG and 
ULIRG luminosities, because project OT1\_dsanders\_1 systematically addressed these. 
Objects below the LIRG threshold enter our sample via the lower luminosity ones
among the RBGS galaxies observed by other proposals, and because some IRAS-defined RBGS 
(U)LIRGs at the better \herschel\ resolution are found to be sub-LIRG pairs or have
companions. Also, other galaxies are serendipitously found within the mapped area. 
For a better 
sampling of normal near-main-sequence galaxies, we also include KINGFISH 
\citep{kennicutt11} galaxies. Since this work focusses on normal star-forming
galaxies and starbursts, we excluded the few dwarfs and low mass systems below
log(M$_*$)=8.5 as well as the elliptical NGC~1404 that are part of the 
KINGFISH sample. Below, we refer to the 
combination of RBGS galaxies, KINGFISH galaxies, and the serendipitous 
galaxies included in our analysis as `galaxies', with a total of
306 galaxies. Our study in addition 
includes Palomar-Green \citep[PG,][]{boroson92} QSOs
from \herschel\ proposals OT1\_lho\_1, OT1\_rmushotz\_1, and OT1\_hnetzer\_1,
\citep[see][]{petric15} 
to which we refer as `QSOs'. We have 93 sources labelled
QSO in our sample, because 2 of the 87 \citet{boroson92} objects 
(PG1226+023 and PG1444+407) were not 
observed by PACS, but 8 QSOs of equivalent properties were additionally
observed in OT1\_lho\_1 or OT1\_hnetzer\_1. To assess the stability 
of the point spread function, we use observations of reference stars 
that were obtained as part of the PACS calibration program.

Several mappers have been successfully used to derive images from the PACS 
data, which are characterized by significant 1/f noise in the individual 
detector
timelines. Since our goal is to derive intrinsic FIR sizes by comparing
the (slightly broadened) observed source FWHM to that of a PSF reference, 
we use the simplemost `masked highpass filtering' strategy which can 
produce a very stable
PSF (see below). In that specific respect it is still superior to mappers 
such as, e.g., UNIMAP \citep{piazzo15} or Scanamorphos \citep{roussel13}. 
While having clear advantages in other respects, these mappers
may sometimes produce residuals near very 
bright sources and/or slightly less stable PSF wings 
\footnote{See `PACS map-making tools: analysis and benchmarking, 30 March 2014' {\tt http://herschel.esac.esa.int/twiki/pub/Public/\\ PacsCalibrationWeb/pacs\_mapmaking\_report14\_v2.pdf}}. 

The reduction follows the standard masked high pass filtering script 
provided with the \herschel\ HIPE software. To obtain 
the most stable PSF, we always use a 20\arcsec\ radius circular patch 
mask to exclude the source from the running median high pass 
filter. We always centered this patch mask on the actual source position.
We start from the \herschel\ science archive pipeline version 13.0 `Level 1' 
data. This is important since version 13.0 for the first time implemented an 
improved reconstruction of the actual \herschel\ pointing, reducing
effects on the PSF by uncorrected pointing jitter. From the maps we measure
circularized Gaussian FWHMs, and aperture
photometry centered on the source that is corrected to total fluxes using 
standard PACS encircled energy fraction curves. By default, we use apertures
of diameter 14\arcsec\ at 70 and 100~\mum\ and 24\arcsec\ at 160~\mum\/. Larger 
apertures are used where required by a clearly extended morphology. We derive
rest frame 40--120~\mum\ luminosities \lfir\ by fitting SED templates from
\citet{berta13} to the PACS fluxes and integrating the best fit template 
over that range.
We use 40--120~\mum\ \lfir\ rather than 8--1000~\mum\ \lir\ because it is less 
AGN contaminated, better sampled by PACS, and one of the common conventions
in the FIR literature going back to IRAS.   
Most sources have been observed with two different
scan angles each in the 70~\mum\ `blue' and 100~\mum\ `green' band, each time
simultaneously with the 160~\mum\ `red' band. The two angles for one 
of the blue or green bands are 
almost always observed in immediate succession, while the other band may be 
obtained at a much later date. To mitigate any slight pointing shifts 
that can arise due to a different star tracker orientation or 
different \herschel\ thermal state between 
these two dates, we average images from the two scan angles before fitting
and deriving fluxes, but average the `red' results from the two epochs only
at the level of the derived FWHM and fluxes. With the exception of the source 
position that is used to center the map and the mask, all other reduction 
parameters are applied identically to all sources.

\subsection{PSF stability}

\begin{figure}
\centering
\includegraphics[width=\hsize]{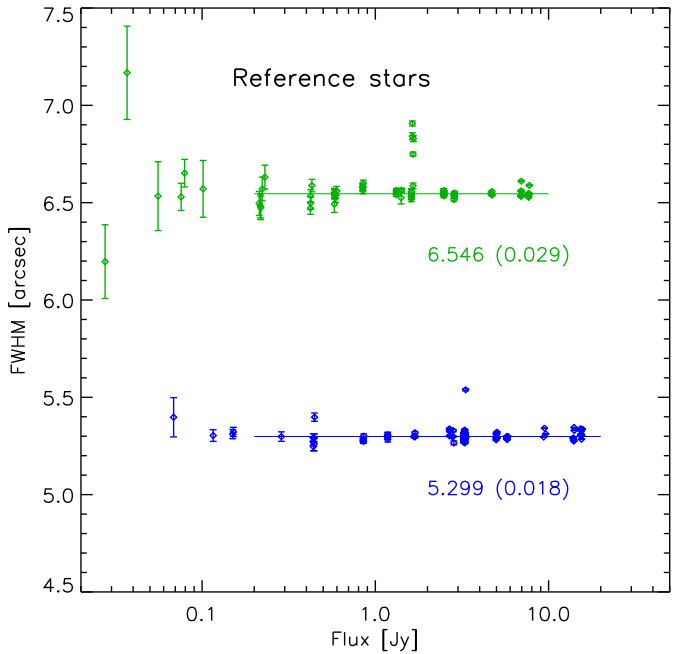}
\caption{Measured FWHM at 70~\mum\ (blue) and 100~\mum\ (green) from 
185 pairs of observations of 17 different reference stars. The resulting
mean FWHM in arcsec and dispersion is indicated, as derived from bright 
sources in the flux range indicated by the horizontal line.}
\label{fig:psfstability}
\end{figure}

To assess the stability of the PSF in our maps, we used the primary and 
secondary flux calibrators 
$\beta$And,
$\alpha$Ari,
$\alpha$Boo,
$\omega$Cap,
$\alpha$Cet, 
$\alpha$CMa, 
$\gamma$Dra, $\delta$Dra,
$\beta$Gem,
$\epsilon$Lep, 
$\beta$Peg,
$\alpha$Tau,
HD41047, HD138265, HD139669, HD159330, HD170693.
These were observed repeatedly (total 185 pairs of observations) as part 
of the PACS calibration program and cover a wide range of fluxes. Data 
were processed and Gaussian fitted exactly as for the science targets. 
Fig.~\ref{fig:psfstability} shows the resulting PSF FWHM for the blue and 
green bands. Results are remarkably stable for bright $\gtrsim$0.2~Jy stars,
and do not show a trend with flux.
The resulting mean FWHM and dispersion are 5.299\arcsec\ 
(dispersion 0.018\arcsec\/) and 6.546\arcsec\ (dispersion 0.029\arcsec\/) 
for the blue and green bands after 
clipping 2 (4) outliers, respectively. These few outliers do not represent 
problematic non-point sources, such as dust shell stars or multiple stars, 
since FWHM values consistent with 
the mean were obtained for the same objects at different epochs.
They rather indicate that in a small $\sim$3\%\ of observations even the 
improved pipeline v13.0 pointing reconstruction has imperfections leading to 
slightly enhanced FWHM. A basic limit to the PSF stability in PACS scan 
observations is set by a random 0-25~msec synchronization jitter between 
instrument and pointing data, set at the start of an onboard control 
procedure. The effect of this jitter for the observing mode that we 
used is below the observed dispersion.  
An equivalent analysis of reference star observations for 
the red band provides circularized FWHM 10.632\arcsec\ 
(dispersion 0.118\arcsec\/). We adopt these values as 
the FWHM and uncertainty of the PSF for a point source with Rayleigh-Jeans 
spectral slope, as reduced with our specific procedure.

\subsection{Effects of source color}

The considerable spectral width of the PACS filters leads to a dependence of 
PSF width on spectral 
shape. This is confirmed on sky
\footnote{{\tt http://herschel.esac.esa.int/twiki/pub/Public/\\ PacsCalibrationWeb/bolopsf\_22.pdf}},
but not accurately
characterized empirically, due to the lack of bright celestial point sources 
with a variety of non-Rayleigh-Jeans slopes. We rather use `as built' 
\herschel\ 
modelled PSFs
\footnote{{\tt http://herschel.esac.esa.int/twiki/pub/Public/\\ PacsCalibrationWeb/PACSPSF\_PICC-ME-TN-029\_v2.0.pdf}} 
for a range of spectral slopes $S_{\nu}\propto\lambda^{\alpha}$, $\alpha=-2\ldots 4$. To account for optical 
imperfections of PACS which are not included in the model PSFs, calibration 
uncertainties, and residual 
uncorrected pointing jitter these model PSFs were Gaussian convolved to 
match the observed stellar PSF FWHM for the case of a Rayleigh-Jeans 
spectral slope. Results imply that between 
$\alpha=-2$ and $\alpha=2$, the PSF FWHM increases by about 2.0\%, 3.4\%, 
5.5\% for the blue, green, red band, respectively.

\subsection{Determination of size of the far-infrared source}

We attempt to measure sizes only for sources with integrated S/N$\geq$10 
in the given photometric band.
To the PACS maps we fit elongated 2d Gaussians using IDL mpfit2dpeak.pro
and a consistent fit area for all sources,
and derive the circularized $FWHM_{\rm obs}=\sqrt{(FWHM_{x}^2+FWHM_{y}^2)/2}$.
We derive the intrinsic source FWHM by subtraction in quadrature 
$FWHM_{\rm source}=\sqrt{(FWHM_{\rm obs}^2-FWHM_{\rm PSF}(\alpha )^2)}$
based on the PSF FWHM derived above and corrected to the spectral slope 
$\alpha$ that is matching the source's PACS photometry. The subtraction 
in quadrature is strictly correct only
for Gaussian structure of both source and PSF. We have verified that the
non-gaussianity of the PSF leads to a systematic overestimate of the intrinsic 
FWHM of a Gaussian 
source by $\lesssim$10\%. We do not attempt to correct for any specific value 
given it would only apply to a strictly Gaussian source structure. The 
error estimate for
$FWHM_{source}$ includes the uncertainty of $FWHM_{obs}$ due to the 
source's S/N, the dispersion of bright reference stars around $FWHM_{PSF}$,
and the effect of the uncertainty in spectral slope. We consider a source as 
resolved if the difference of $FWHM_{\rm source}$ and $FWHM_{PSF}$ is larger
than three times this error estimate. For brevity, we denote 
below with \reblue\ the intrinsic half light radius at 70~\mum\ ($=0.5\times FWHM$ 
for a Gaussian source), after conversion 
to a physical scale in kpc based on distances from redshift and our 
adopted cosmology, with the exception of redshift-independent distances
for the KINGFISH sources which were taken from \citet{kennicutt11}. 
\regreen\ and \rered\ analogously denote the physical half light radii at 
the longer PACS wavelengths. 

\subsection{Resolved sources}
The reduction described previously is optimized for sources with intrinsic
width similar to or below the width of the PSF. This includes all the QSO 
hosts. For galaxies that are resolved over several beam widths, as 
frequently found
among the nearby galaxies and less luminous IR selected objects, the specific 
masking/filtering strategy reduces the outermost emission. For objects
that were found to be clearly extended on inspection of PACS maps and/or 
with an initial 
measured 70~\mum\ FWHM above 12\arcsec\/, we hence reverted to the V13.0 
pipeline reduced `Level 2.5' JScanam \citep{graciacarpio15} maps, or to 
the publicly released KINGFISH DR3 Scanamorphos maps. Intrinsic sizes 
were again derived by Gaussian
fits and subtraction in quadrature of the circularized width. 
Both reductions give consistent results at intermediate source sizes.   

\subsection{Double or complex sources}

Discussing physics of galaxies on the basis of sizes derived in this way is 
only meaningful if the size refers to a single galaxy, and not galaxy
pairs or interacting systems. This is of course 
acknowledging that even in this case the size represents an overall 
scale of the galaxy's FIR emission, that may encompass significant 
substructure.
The measured size is not meaningful for interacting systems that have 
a separation of the components of order the PSF width, and down to 
the scales accessible to the 
method. The measurement then mostly reflects distance and relative strength of
FIR emission of the two galaxies rather than the structure of a galaxy. We 
address this issue the following way:
\begin{itemize}
\item Wide $\gtrsim$1\arcmin\ double systems with well separated galaxies 
and redshifts available for both were processed separately for both 
components, and both galaxies are used in the analysis.
\item Wide and $\sim$0.5\arcmin\ intermediate doubles where galaxies overlap
and/or high pass filtering affects the structure of the respective other 
component were 
discarded. These would need mapping and fitting strategies beyond the scope
of this work, in order to measure high accuracy FIR sizes. A few intermediate 
scale doubles were kept if the secondary component was clearly below 1/10 of 
the primary component's flux and not affecting the width measurement.    
\item Sizes were measured for single and close $\lesssim$ 10\arcsec\ double
sources. However, where inspection of multiwavelength data accessible
via NED and the literature \citep[e.g.,][]
{bahcall97,scoville00,guyon06,haan11,kim13,petty14,surace01,veilleux06,veilleux09} 
indicated that the source is in fact a close
double/interacting system, the source was flagged even if a double nature
is not evident in the \herschel\ images. These flagged sources are not used 
in the analysis below, unless explicitly mentioned.  
\end{itemize}
In summary, size measurements are obtained for single galaxies (including 
advanced mergers) and for widely separated interacting systems, but disfavour
close interacting systems with separation of order the galaxy size, and where
both components emit in the FIR. Our size measurements are summarized in 
Table~\ref{tab:longtable}.

\section{Results}
\begin{figure}
\centering
\includegraphics[width=\hsize]{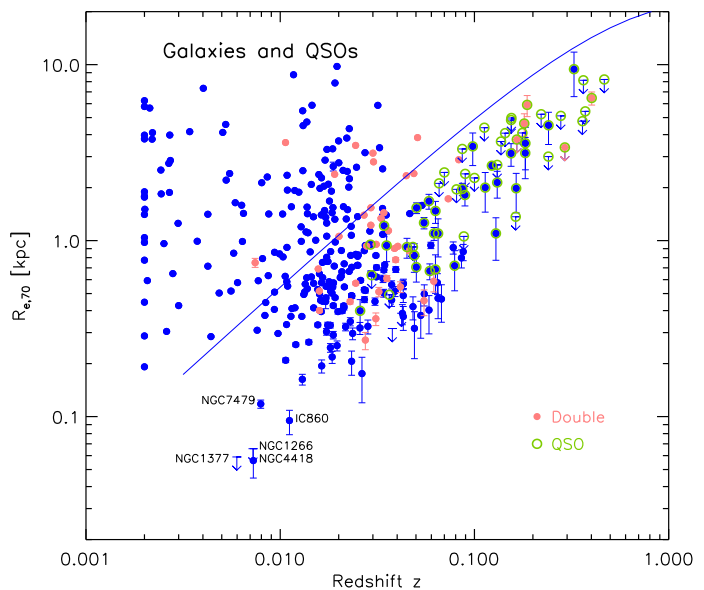}
\caption{Intrinsic half light radii of galaxies and QSOs
in the 70~\mum\ blue band as a function of redshift. Symbols or limit for 
QSOs are encircled in green. The blue line 
indicates the physical size corresponding to the half width at half maximum 
of the PACS point spread 
function at that wavelength. Pink symbols mark results for close 
doubles/pairs.
These are not used below unless explicitly mentioned. The most local galaxies 
have been placed at z=0.002 to limit plot size. A number of galaxies with
very compact FIR regions are identified. For many bright and large targets, 
error bars are smaller than the symbol size. Lower redshift galaxies with
typically higher S/N can be resolved to smaller fractions of the PSF than the
fainter QSOs.}
\label{fig:bluesizeredshift}
\end{figure}

\begin{figure}
\centering
\includegraphics[width=\hsize]{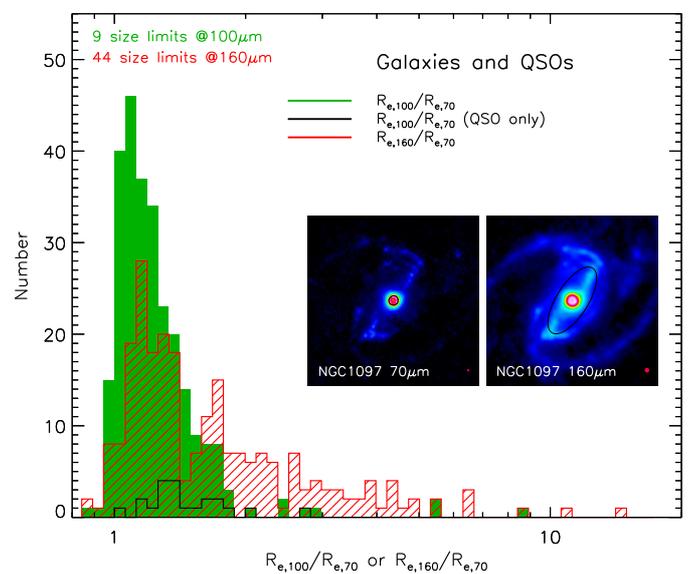}
\caption{Histograms of the ratio of the intrinsic half light radius in the
100~\mum\ (160~\mum\/) band to the half light radius in the 70$\mu$m band, for
sources with significant measurements in either band. The numbers
of sources with a measurement at 70~\mum\ but only a limit at the longer 
wavelength are listed in the top left, these will typically have size 
ratios $\sim$1--2. The case of NGC~1097 (insets) visualizes how a warm
central starburst superposed on a cold large disk can cause large ratios
\rered\//\reblue\ from fitting single Gaussians. Overplotted to 
the inset for each band are the ellipses at half maximum of the 
elongated Gaussian fit (black) and the beam (red, bottom right).}
\label{fig:bluegreenredsizes}
\end{figure}

Fig.~\ref{fig:bluesizeredshift} shows the derived intrinsic half light radii
\reblue\ (effective radii) 
in the 70~\mum\ blue band as a function of redshift. The 355 of 399 
sources that have a photometric S/N $>10$ in that band are plotted. The 
blue line indicates the physical size corresponding to the half width at 
half maximum of the PSF. In comparison to the data, this suggests that 
in cases with favourably high S/N the broadening by a source with intrinsic 
Gaussian FWHM about one fifth of the PSF FWHM is still detectable and leads 
to a size measurement, i.e. the difference of source and PSF size is at least
three times the error.  For this sample, almost all
z$<$0.1 results are actual measurements and upper limits only become common
for the sources at z$>$0.1, all from the QSO sample. Limits in the 
plot are placed at twice the error plus the larger of 
the nominal measurement and zero. There is 
a two order of magnitude spread in the measured intrinsic FIR sizes of 
the non-QSO targets. Sect.~\ref{sect:scalings} discusses scalings with 
basic quantities that are contributing to this spread.

Signal-to-noise and PSF width lead to the largest fraction of significant
size measurements being obtained in the 70~\mum\ blue band. It is 
81\% compared to 80\% at 100~\mum\
and 71\% at 160~\mum\/, where the remaining objects include systems with
both upper limits and photometric S/N$<$10 in a band, which were not 
probed for size. We focus our analysis on 70~\mum\ because of that, and 
because in star-forming galaxies 70~\mum\ emission is more tightly linked to
SFR than longer far-infrared wavelengths. \citep[e.g.,][]{calzetti10}. 

\begin{figure*}
\centering
\includegraphics[width=0.49\hsize]{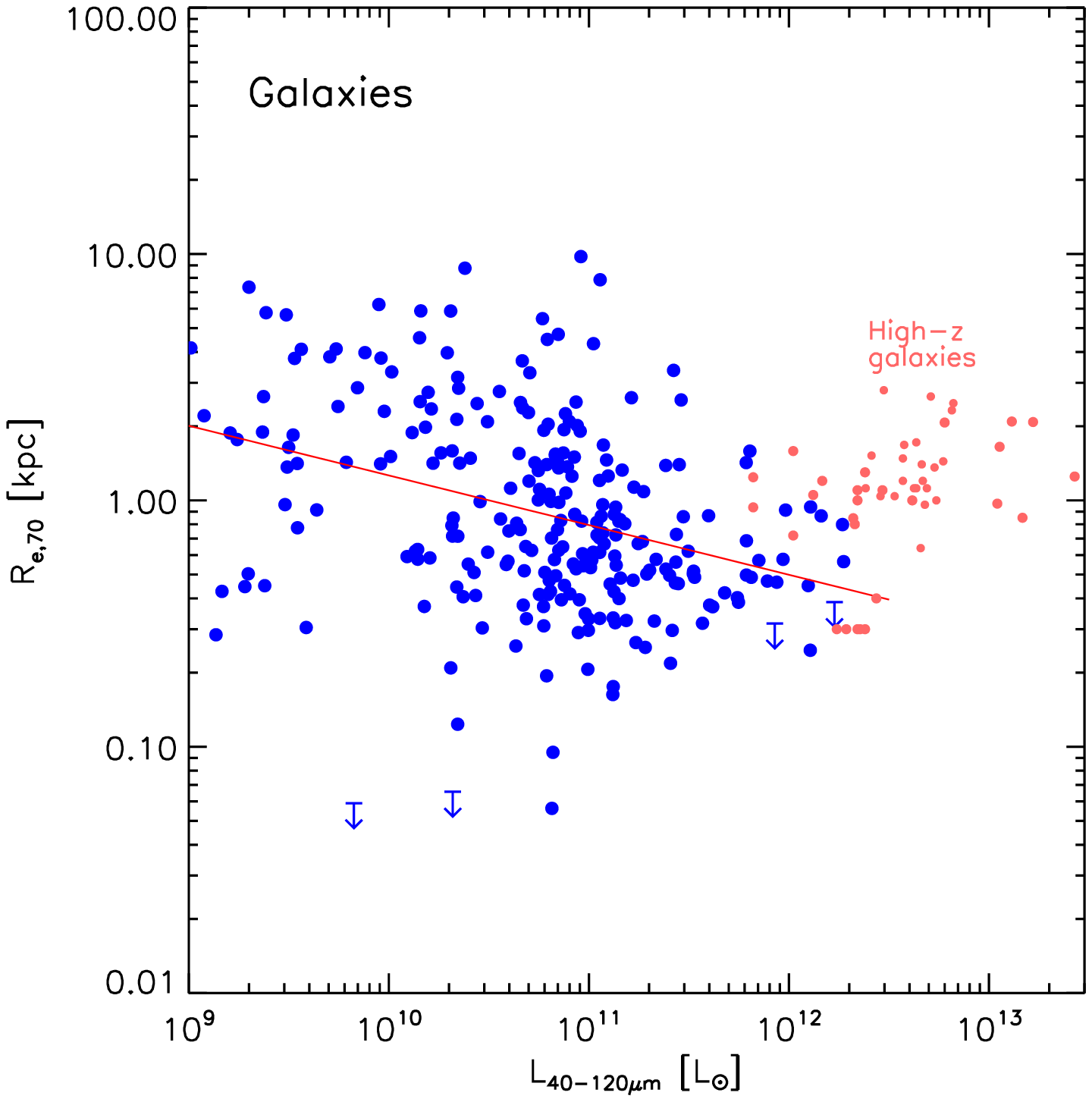}
\includegraphics[width=0.49\hsize]{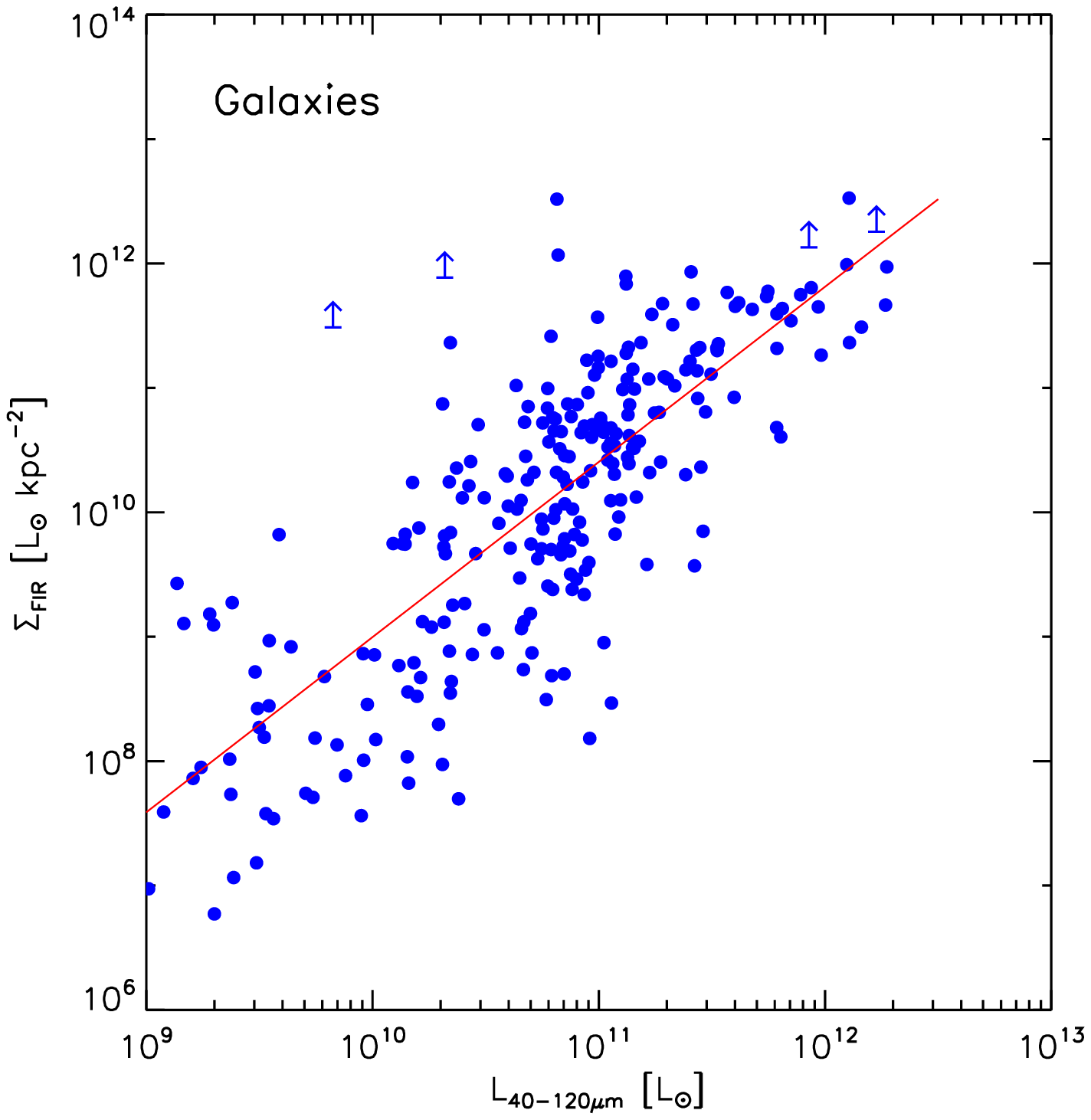}
\caption{Scalings of half light radius at 70~\mum\ (left) and of corresponding
rest frame FIR surface brightness (right) with far-infrared luminosity.
251 galaxies with \lfir$>10^9$\lsun\ are included to avoid dwarfs and 
faint companions. The red line is a power law fit. In the left panel, 
sizes derived
in the literature for high redshift submm galaxies at longer rest wavelengths 
$\sim$150--400~\mum\ are included for comparison (pink symbols). These were 
not used for the fit.}
\label{fig:lfirscalings}
\end{figure*}

Still, a
comparison between sizes for different wavelengths in the same sources is
in place. Fig.~\ref{fig:bluegreenredsizes} shows size ratios for the
green/blue and red/blue bands, for sources with a significant size 
measurement in both bands. Many sources show slightly larger sizes at the 
longer wavelength, but still close to the 70~\mum\ size. This is plausible
given the typically negative outward dust temperature gradients in galaxies
\citep[e.g.,][]{hunt15}. Sources with a 70~\mum\ size measurement but unresolved
at the longer wavelengths with their larger PSF will also be consistent 
with such size ratios near 1. 
There is however a significant tail to the distribution, up to size ratio
\regreen\//\reblue$\sim$5 in green vs. blue and \rered\//\reblue$\sim$10 in 
red vs. blue. Since many of these objects with large size ratios
are well resolved nearby targets, visual inspection of the maps helps to
identify the cause. They
typically show a warm central source, e.g., a circumnuclear starburst, 
superposed on a larger disk with star-forming regions and with diffuse
emission that is more prominent at 160~\mum\/. A single component Gaussian fit 
may then 
emphasize the central component at 70~\mum\/, and the disk at 160~\mum\/.
A similar argument as for a circumnuclear starburst applies if there is
AGN-heated near-nuclear dust. 
A clear example for the effect of a central starburst  is NGC~1097,
where the 70~\mum\ fit gives FWHM$\sim$20\arcsec\/, driven by the well-known 
prominent r$\sim$900~pc circumnuclear starburst ring 
\citep[e.g.,][]{sandstrom10}, while the 160~\mum\
fit returns FWHM $\sim$2\arcmin\ because of the relatively stronger emission 
from the diffuse disk (see also inset to Fig.~\ref{fig:bluegreenredsizes}). 
This is obviously a consequence of fitting a one 
component Gaussian in a situation where a two component fit with a wavelength
dependent weight would be more appropriate. For consistency with the analysis of the majority
of targets that are somewhat broadened compared to the PSF, but not 
well resolved, and to typical high-z observations, we adopt single 
component fits in the following discussion. We use 70~\mum\ because of the 
smallest PSF and the closer link of emission at this wavelength to SFR for 
star-forming galaxies.

Size differences in the PACS bands up to a factor of few warrant a word
of caution towards the interpretation of interferometric 
(sub)mm continuum images of high redshift galaxies. With few exceptions, 
they are taken at rest
wavelengths clearly longer than 160~\mum\/, and may often not have
the quality for well resolved maps or complex models. The measured sizes may 
hence 
overestimate the size of the star-forming region that is of interest for, e.g.,
studies of the Kennicutt-Schmidt relation. Continuum fluxes on the
Rayleigh-Jeans tail of the SED are proportional to dust mass and 
(linearly) to dust temperature. Such long wavelength interferometric
maps hence more directly trace the dust surface density rather than the
SFR surface density. If the dust properties and dust-to-gas ratio 
are well constrained, these long wavelength continuum fluxes are a better 
probe of the gas surface density than the SF surface density in the 
Kennicutt-Schmidt relation.
Of course, gas-rich high-z systems may have different structure
and {\em quantitative} size ratios between different rest wavelengths than 
local galaxies of similar SFR. For example, high-z galaxies with intense 
star formation throughout the disks \citep[e.g.,][]{foerster09,nelson15} 
will almost certainly show lower ratios between size at longer wavelengths
and size at rest 70~\mum\ than the extreme values in a few 
local LIRGs with a quite passive disk and a bright circumnuclear
starburst -- these may in fact be a poor analog to high redshift galaxies 
of similar SFR. Even if sizes likely differ by smaller factors, the caveat is 
relevant at high redshift as it is locally.

\section{Discussion}
\subsection{Scalings of FIR source size and surface
brightness with FIR luminosity, main-sequence offset, and 
FIR color}
\label{sect:scalings}

\begin{figure*}
\centering
\includegraphics[width=0.49\hsize]{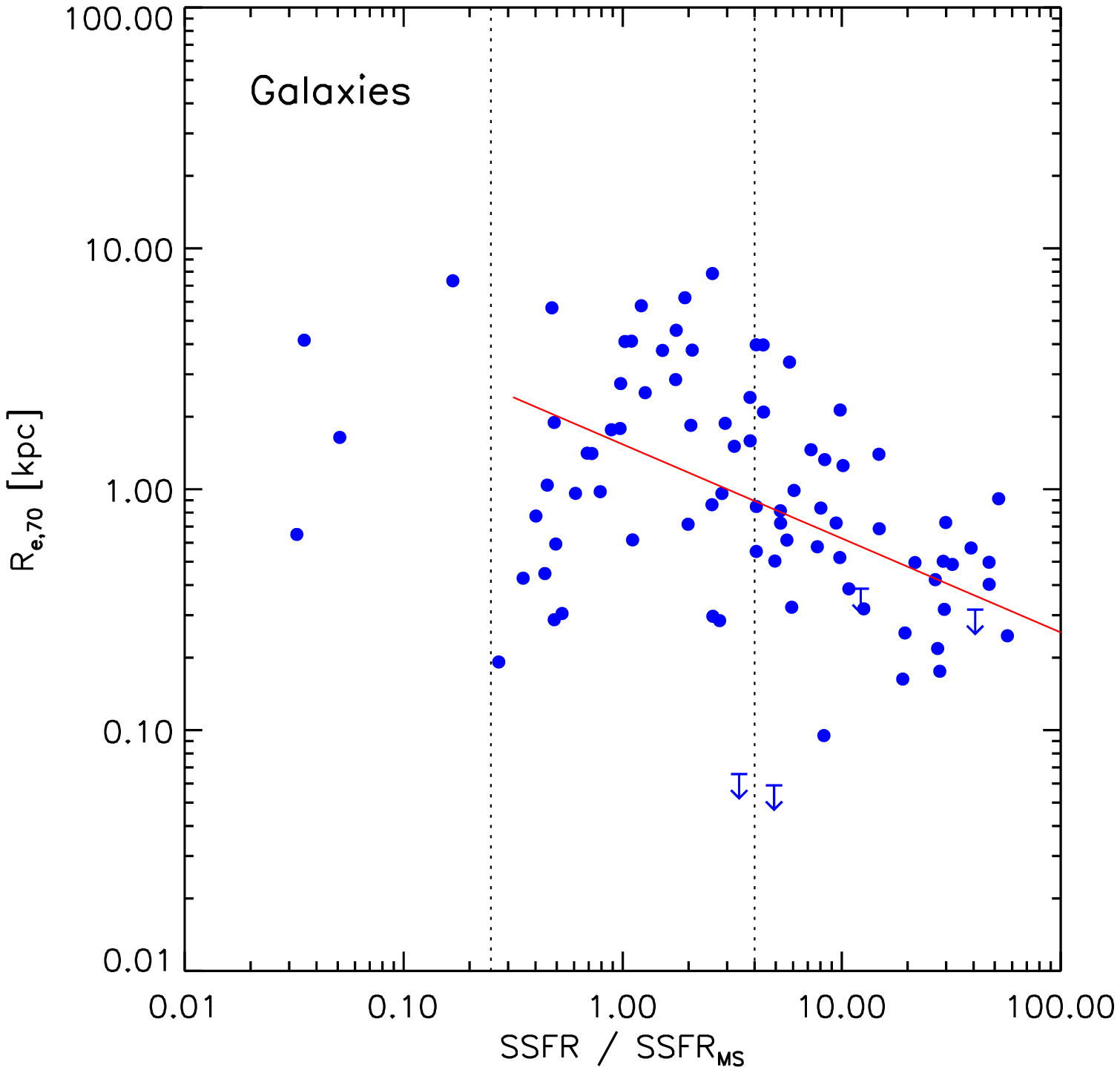}
\includegraphics[width=0.49\hsize]{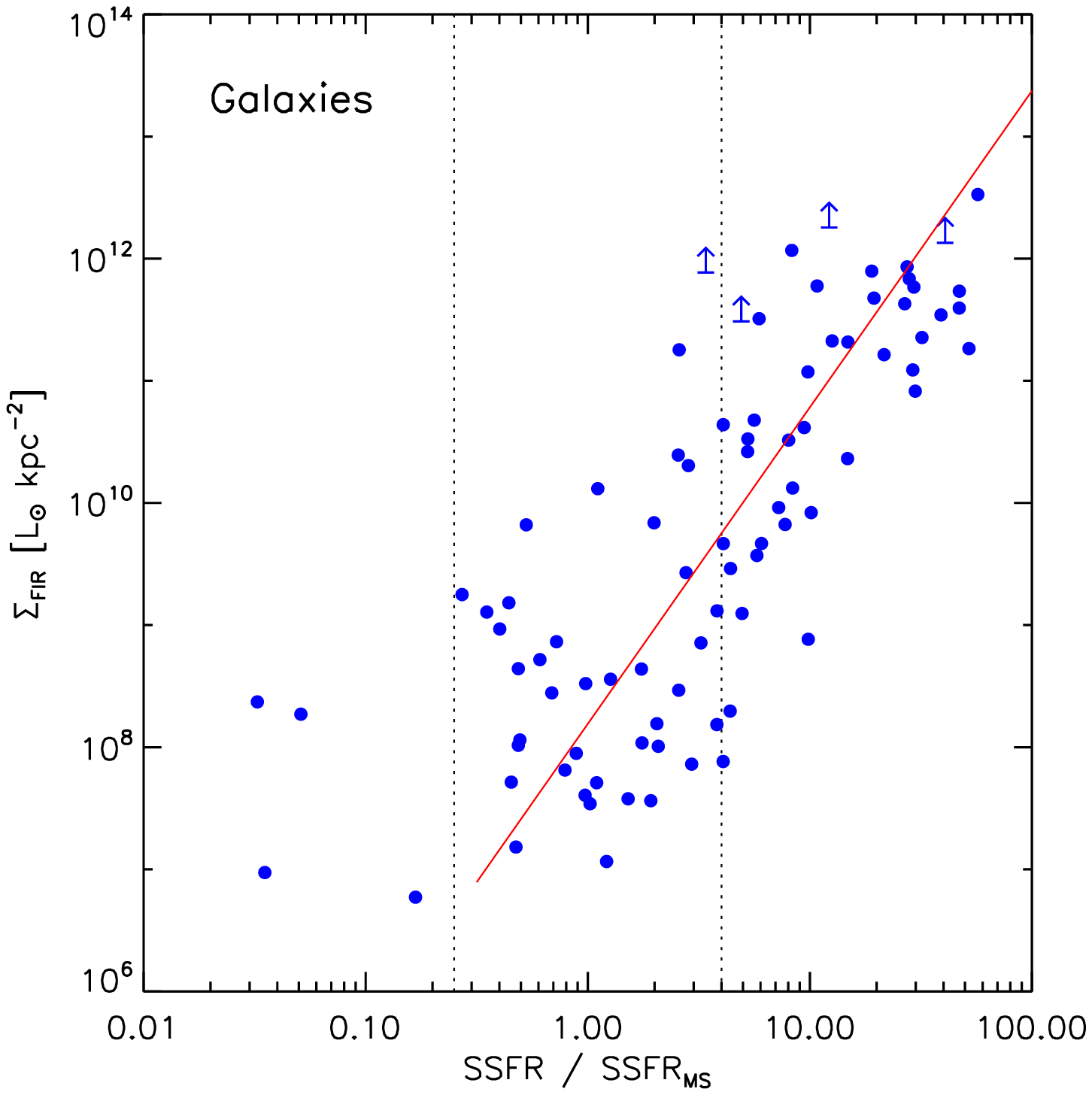}
\caption{Scalings of FWHM size at 70~\mum\ (left) and of corresponding
rest frame FIR surface brightness (right) with distance from the 
\citet{renzini15} star-forming main-sequence. The figure includes
86 galaxies with published stellar masses.
The red line is a power law fit. Vertical dotted lines indicate a range of
$\pm$0.6~dex around the main-sequence, as often used to separate main-sequence
galaxies from bursting or passive objects.}
\label{fig:dssfrmsscalings}
\end{figure*}

The very large $\sim$2~dex spread in size of the FIR emitting regions
motivates an investigation of possible links with basic galaxy parameters. 
We start with the 40-120~\mum\ luminosity \lfir\/. Fig.~\ref{fig:lfirscalings} 
left indicates a decrease of size with increasing FIR luminosity. Using
mpfitexy.pro \citep{williams10} to fit a log-linear relation (power law) we 
obtain:
\begin{equation}
log (R_{\rm e,70}) = (0.101\pm 0.036) - (0.202\pm 0.035)\times (log(L_{\rm FIR})-10).
\label{eq:size_fir}
\end{equation}

We may also investigate the far-infrared surface brightness 
$
\Sigma_{\rm FIR}\equiv L_{\rm FIR}/2\pi R_{\rm e}^2
$,
noting that for a Gaussian source the half light radius $R_{\rm e}$ 
equals to FWHM/2. Fig.~\ref{fig:lfirscalings} right shows the strong increase
of surface brightness with \lfir\/, which is fit by
\begin{equation}
\begin{split}
log (\Sigma_{\rm FIR}) & = &  \pushleft{(8.997\pm 0.072)} \\
                       &   & + (1.408\pm 0.071)\times (log(L_{\rm FIR})-10)
\end{split}
\label{eq:sigfir_fir}
\end{equation}
and
\begin{equation}
log (\Sigma_{\rm SFR})  =  -1.117 + 1.408\times log({\rm SFR}),
\label{eq:sigsfr_sfr}
\end{equation}
where the second equation for the surface density of SFR 
(\msun\ yr$^{-1}$ kpc$^{-2}$) uses our adopted linear conversion 
${\rm SFR}=1.9\times 10^{-10}$\lfir\/. Uncertainties are hence the same 
as for $\Sigma_{\rm FIR}$.
Both $\Sigma_{\rm FIR}$ and $\Sigma_{\rm SFR}$ refer to the plane of the sky. 
We did not attempt to correct for inclination, given that interacting systems
can have complex morphologies and/or the inclination is not constrained for
the FIR emitting regions.
The increase of far-infrared surface brightness and SFR surface
density with \lfir\ is clearly super-linear, due to a combination of the 
trivial increase with luminosity and the decreasing size.

We also investigate the relation of size \reblue\ and surface brightness 
\sigfir\ to
the distance of a galaxy from the star-forming main-sequence. Here we adopt 
as location of the main-sequence in the local universe 
$log({\rm SFR})=0.76\times log(M_{\ast})-7.64$
\citep{renzini15} and its equivalent main-sequence specific star formation 
rate ${\rm sSFR}_{\rm MS}$.
Stellar masses for part of our sample are available from
\citet{kennicutt11} and from \citet{u12}. Concerning the U et al. stellar 
masses we adopt $log(M_{\rm fit})_{\rm Cha}$ from their Table~10 and exclude 
objects where their
stellar mass combines two galaxies that we separate with \herschel\/.
For our sample selected to contain many (U)LIRGs, 50\%\ of the sources
with a stellar mass assigned have $log({\rm sSFR}/{\rm sSFR}_{\rm MS})>0.6$, 
a commonly adopted division between main-sequence galaxies and 
starbursting outliers above the main-sequence \citep{rodighiero11}. 
Fig.~\ref{fig:dssfrmsscalings} shows the trends for \reblue\ and \sigfir\ 
with distance from the main-sequence. 
Excluding a few semi-passive KINGFISH objects at 
$log({\rm sSFR}/{\rm sSFR}_{\rm MS})<-0.5$ we obtain:
 \begin{equation}
\begin{split}
log (R_{\rm e,70}) & = &\pushleft{(0.19\pm 0.07)} \\
                   &   &  - (0.39\pm 0.08)\times log({\rm sSFR}/{\rm sSFR}_{\rm MS}),
\end{split}
\end{equation}
\begin{equation}
\begin{split}
log (\Sigma_{\rm FIR}) & = &\pushleft{(8.19\pm 0.17)}\\
                       &   & +(2.59\pm 0.20)\times log({\rm sSFR}/{\rm sSFR}_{\rm MS}),
\end{split}
\label{eq:sigfir_dssfrms}
\end{equation}
\begin{equation}
log (\Sigma_{\rm SFR}) = -1.53 + 2.59\times log({\rm sSFR}/{\rm sSFR}_{\rm MS}).
\label{eq:sigsfr_dssfrms}
\end{equation}

Correlations also exist between \reblue\ or \sigfir\ and  sSFR. This 
is expected from
the trends shown with sSFR/sSFR$_{\rm MS}$, given that the adopted  
main-sequence slope is not too far from one in SFR vs \mstellar\/, i.e. quite 
flat in sSFR. The IR selected (U)LIRGs
reach up to sSFR$\gtrsim 10^{-9}$yr$^{-1}$ and
$\Sigma_{\rm SFR}\sim$100~\msun\/yr$^{-1}$kpc$^{-2}$, well above the range
of the local universe normal and passive galaxies for which \citet{wuyts11b}
used optical diagnostics to study the relation of sSFR and 
surface density of star formation. A trend between (total) infrared surface brightness
and sSFR for local galaxies was also noted by \citet{elbaz11} (their Fig.~16),
with some outliers possibly due to the use of mid-IR based sizes.
When relating \reblue\ and \sigfir\ to stellar mass, the scatter is large 
and no strong trends are discernible.
  
\begin{figure*}
\centering
\includegraphics[width=0.49\hsize]{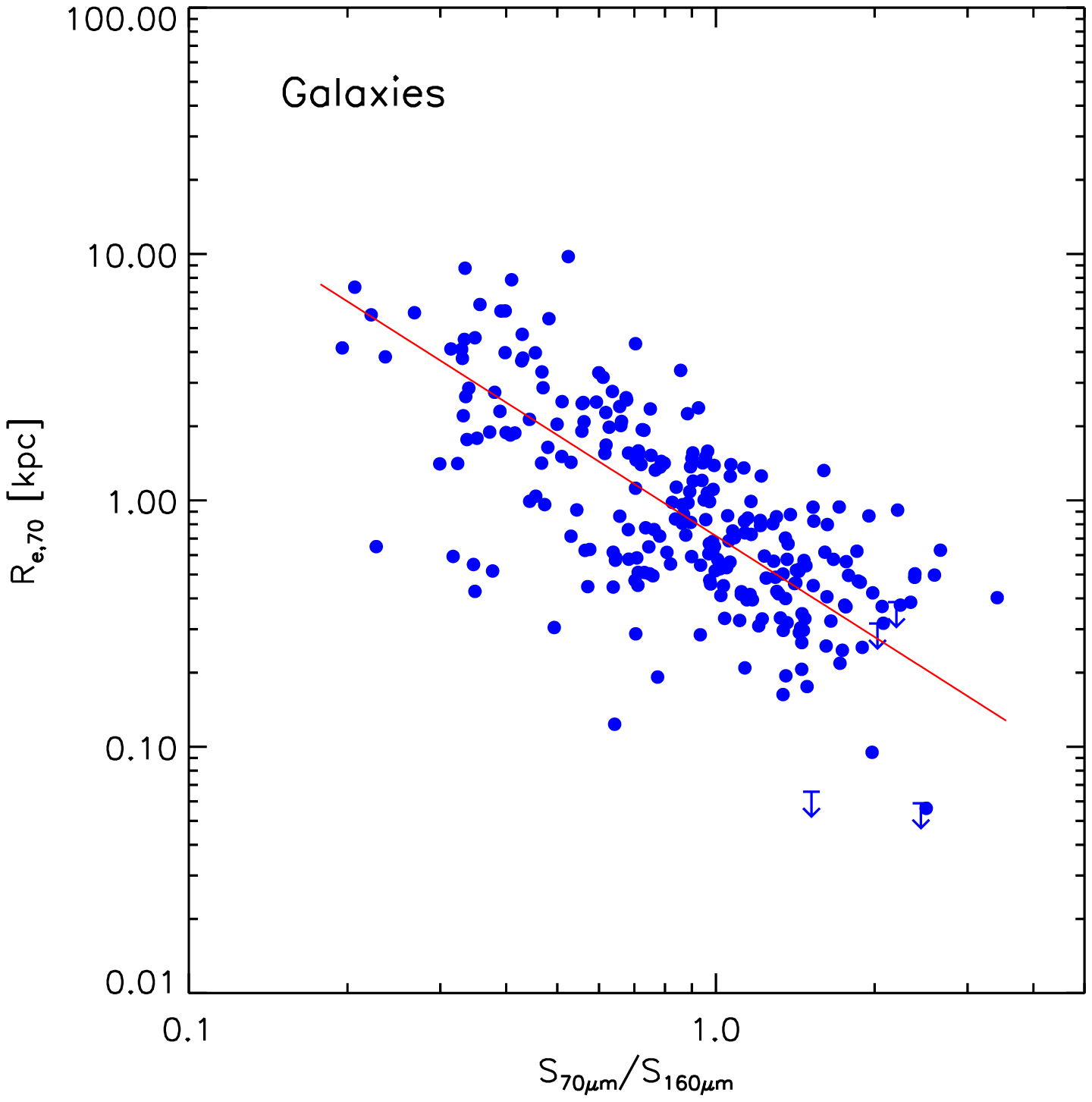}
\includegraphics[width=0.49\hsize]{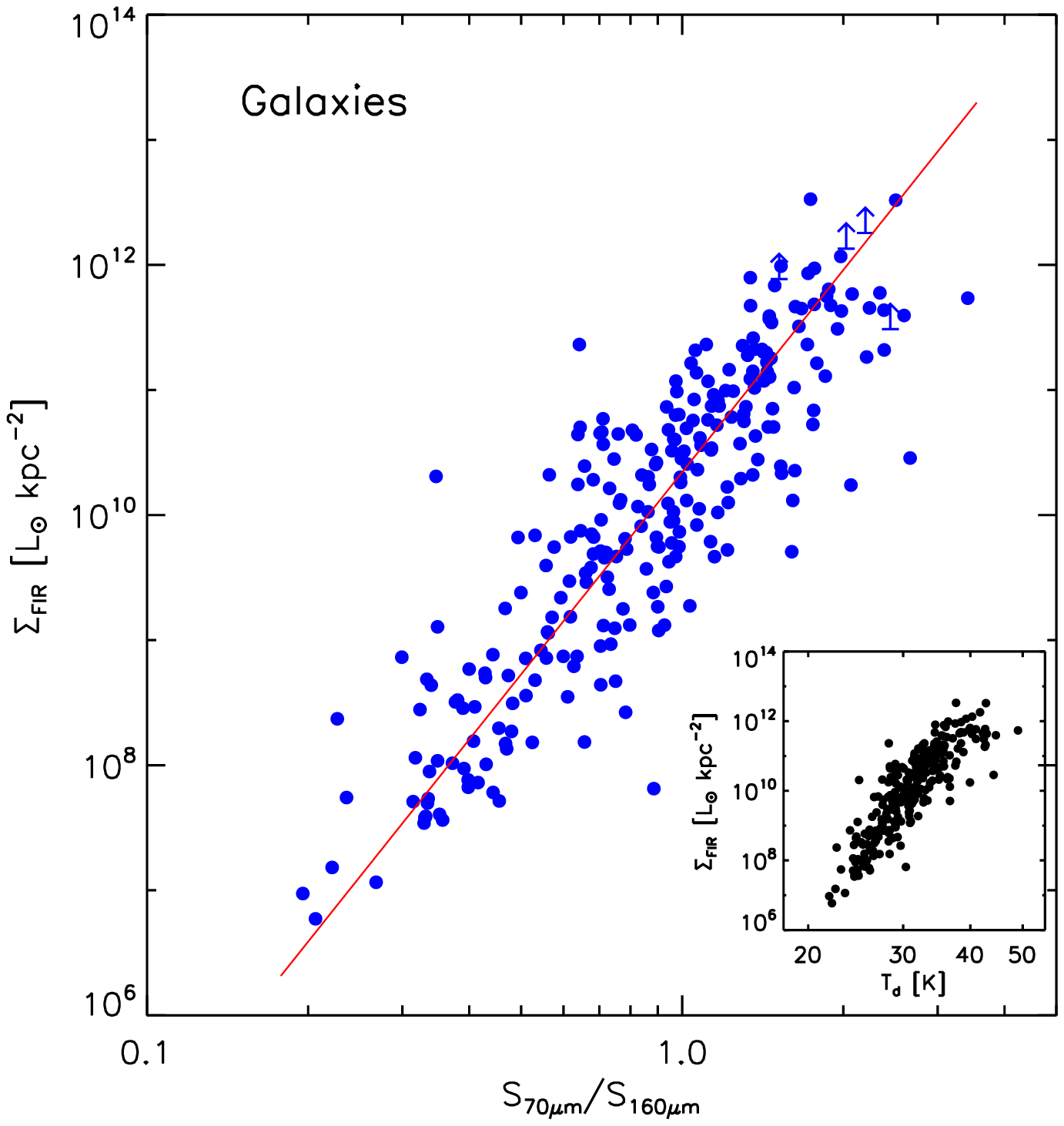}
\caption{Scalings of FWHM size at 70~\mum\ (left) and of corresponding
rest frame FIR surface brightness (right) with PACS 70 to 160~\mum\ color,
for 260 galaxies.
The inset to the right panel shows the relation as a function
of dust temperature.}
\label{fig:colorscalings}
\end{figure*}

Finally, relations exist between far-infrared color and size or 
surface brightness of the FIR emission. 
Fig.~\ref{fig:colorscalings} shows very 
clear trends of both quantities with the PACS color $S_{70}/S_{160}$.
In particular the relation with \sigfir\ is remarkable, and takes the form:
\begin{equation}
log (\Sigma_{\rm FIR}) = (10.34\pm 0.04) + (5.37\pm 0.18)\times log(S_{70}/S_{160}),
\end{equation}
\begin{equation}
log (\Sigma_{\rm SFR}) = 0.62 + 5.37\times log(S_{70}/S_{160}),
\end{equation}
 while the relation of size and color is
\begin{equation}
log (R_{\rm e,70}) = (-0.145\pm 0.019) - (1.362\pm 0.076)\times log(S_{70}/S_{160}).
\end{equation}

A relation between \sigfir\ and FIR color, which traces dust temperature, is 
certainly plausible and expected, but it is quite tight with a scatter of 0.6~dex
in \sigfir\ around a relation that covers more than five orders of magnitude
in that quantity. This is notable because the relation compares a local
quantity (the local radiation field intensity determining the
dust temperature) with a surface brightness that is determined solely from
global quantities of the galaxy (\lfir\ and FIR size). In contrast, the 
radiation field intensity to which the FIR emitting grains are exposed
is influenced by the small scale distribution of stars and dust in star 
forming complexes as well as to the density with which sources of 
radiation are distributed over the total volume of the galaxy.
In an extreme toy model, where one simply
places more and more independent star-forming regions with identical FIR 
properties in the same area to drive up \sigsfr\ and \sigfir\/, no such 
relation would exist.

It is also instructive to express the \sigfir\ as function of dust temperature,
where we have converted color to temperature by fitting \citet{dale02} 
SED templates at the source redshift to the color, and using the conversion of 
\citet{magnelli14} to assign dust temperatures to these templates. A 
similar result is obtained for
a  simple $\beta=1.5$ modified blackbody. The result is shown in the inset to
Fig.~\ref{fig:colorscalings} right and can be fit as  
\begin{equation}
\label{eq:sigfir_td}
log (\Sigma_{\rm FIR}) = -(22.5\pm 1.8) + (21.8\pm 1.2)\times log(T_{\rm d}).\\
\end{equation}

This scaling can be compared to expectations from scalings of other quantities,
as found above and as used in the current literature. If we assume
 (1) a relation of far-infrared surface brightness and main-sequence offset 
$log(\Sigma_{FIR}) = \alpha\times log({\rm sSFR}/{\rm sSFR}_{MS}) + c1$ 
with $\alpha$=2.59 and c1=8.19 (Eq.~\ref{eq:sigfir_dssfrms} above), 
(2) that gas depletion time in yr$^{-1}$ scales with main-sequence offset as
$log(M_{\rm gas}/{\rm SFR}) = \xi\times log({\rm sSFR}/{\rm sSFR}_{\rm MS}) + 
c2$ with $\xi$=-0.49 and c2=9.1 \citep[][for z=0 and log(\mstellar )=10.5]{genzel15},
(3) assume a dust-to-gas ratio 100 
\citep[][for solar metallicity]{leroy11}, and  
(4) that the dust is an  optically thin calorimeter with  
$log(SFR/M_{\rm d}) = (4+\beta)\times log(T_{\rm d}) +c3$ with a 
dust emissivity index $\beta=1.5$ and c3=-15.08 
\citep[e.g.,][]{magnelli14,genzel15}, then we expect
\begin{equation}
log(\Sigma_{\rm FIR}) = -\frac{\alpha}{\xi} (4+\beta) log(T_{\rm d}) + 
const = 29.1\times log(T_{\rm d})-34.0, 
\end{equation}
which for the relevant temperature range is in reasonable agreement with 
what was found in Eq.~\ref{eq:sigfir_td},
 given the uncertainties of the various parameters,
and the simplifications in the assumptions made. 
That means, the above set of
scalings and assumptions remains consistent when bringing in size information.
 
Given the data and scatter, we did not attempt to fit more complex 
functional forms than the simple power law (log-linear) relations quoted 
above and summarized in Table~\ref{tab:scalings}. One result from 
comparing \reblue\ and \lfir\ 
(Fig.~\ref{fig:lfirscalings} left) is however worth mentioning.
All ULIRGs are found at similar and relatively small $\sim$0.5~kpc half
light radii, 
but for less luminous $log(L_{\rm FIR})=10\ldots 11$ objects the variation is 
a substantial $\gtrsim$2~dex, ranging 
from the typical  $\sim$1~kpc up to $\sim$10~kpc and down to a few extremely 
compact $\lesssim$100~pc circumnuclear sources. Very small sizes for a ULIRG
far-infrared emitting region are excluded by optical depth arguments 
(Sect.~\ref{sect:opticaldepth}), but the absence of large $\gtrsim$5~kpc FIR 
sources in ULIRGs reflects a true absence of such extended ULIRGs. Similar
findings are made in the mid-IR continuum \citep{soifer00,diazsantos11} and
radio \citep{condon91}. The far-infrared data 
reinforce this with a closer link to the SFR and much reduced risk of a result
that is biased by a compact AGN-heated source. A 
\lfir $\sim 1\times10^{11}$\lsun\ LIRG can correspond to a 10~kpc scale large 
disk, a smaller few kpc region in a (perhaps interacting) galaxy, or even
a compact $\sim$100~pc circumnuclear burst. In contrast, the $\sim$0.5~kpc 
radii
consistently found for ULIRGs are in line with the notion that for the modest 
gas content of galaxies in the local universe, the corresponding huge SFRs
can only be reached by compressing gas and triggering a burst of star 
formation, in an interaction or merging event.

The structural properties of local galaxies are an important comparison
to the structure of evolving galaxies at high redshift.  
\alma\ starts observing samples of high-z galaxies at
longer submm wavelengths, for which the \herschel\ data are the key comparison
despite their somewhat shorter rest wavelengths. Fig.~\ref{fig:lfirscalings} 
left includes some first \alma\ sizes of submm 
galaxies, as measured by \citet{ikarashi15} and \citet{simpson15b} at rest 
wavelengths about 150 to 400~\mum . This comparison
is preliminary since their sources lack individual spectroscopic redshifts 
and determinations
of the SED peak, which makes the adopted individual \lfir\ values uncertain
for the quite diverse SMG population \citep{magnelli12}. We adopted the 
infrared luminosities of \citet{ikarashi15} and estimated them for the 
\citet{simpson15b} sources using the average $S_{850}/L_{\rm IR}$ scaling of 
\citet{magnelli12} at the typical Simpson et al. submm flux. Specifically,
we adopt $L_{\rm FIR}=5.6\times 10^{11} S_{850}$ which also includes
$L_{\rm IR} = 1.5 \times L_{FIR}$ as recommended by \citet{magnelli12}
for such SMG luminosities. We also include
a number of individual SMG continuum size measurements from the literature 
\citep{younger08,younger10,riechers13,riechers14,neri14}, in 
some cases updating
the original luminosity estimates with \herschel\ results 
\citep{magnelli12,smolcic15}. We have also included the sample of 
\citet{harrison16} which consists of five modest luminosity X-ray AGN 
and another galaxy. We group them with galaxies since at 
\lfir\//$L_{\rm Bol,AGN}\sim$10, their AGN are energetically much less 
prominent than in the local QSOs studied in Sect.~\ref{sect:qsosize}. 
This first 
comparison suggests that on average these SMGs, likely at z$\gtrsim$2 and 
in some cases reaching up to z$\sim$6, have roughly twice
the size of local galaxies of similar infrared luminosity. This is consistent
with the notion that massive gas rich high redshift galaxies can reach 
SFRs in excess of 100~\msun\ yr$^{-1}$ in large disks 
\citep[e.g.,][]{foerster09,nelson15}. Because of their increased
gas fractions \citep[e.g.,][]{daddi10,tacconi10,tacconi13,saintonge13,
carilli13}, they can reach high SFRs without the strong gas compression
that is needed to create a local ULIRG. To fully understand the role of 
true structural differences, and the possible dependence of size on wavelength, 
it will be necessary
to compare the local results to (sub)mm studies of high-z samples with 
individual redshifts and reliable derivations of \lir\ and/or SFR, and to rest
frame optical structure \citep[e.g.,][]{vanderwel14}. Submm selection with
its biases on, e.g., dust temperature will have to be supplemented by SFR 
selected and stellar 
mass selected samples covering a wider range of properties than the high-z 
systems in Fig.~\ref{fig:lfirscalings}.

\subsection{Optical depth of FIR emission} 
\label{sect:opticaldepth}

\begin{figure}
\centering
\includegraphics[width=\hsize]{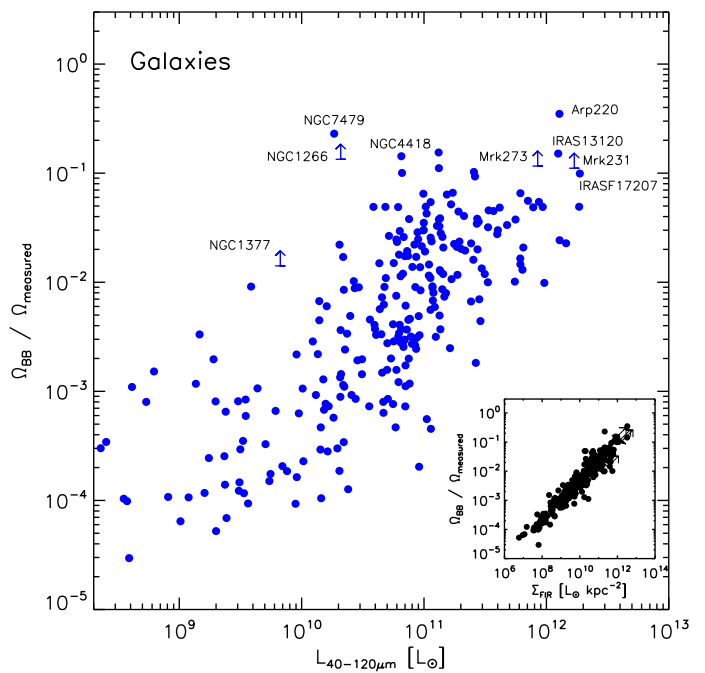}
\caption{Ratio between the solid angle of a blackbody emitting
half of the PACS fluxes at the blackbody temperature fitting those fluxes, and
the measured half light solid angle, for 260 galaxies. 
Values approaching one indicate emission that
is on average optically thick in the far infrared over the measured source 
size. Low values indicate optically thin emission and/or a scattered
distribution of small sources which might still have considerable far-infrared 
optical depth. The inset shows the relation between this ratio of solid 
angles and the far-infrared surface brightness. The tightness is somewhat 
trivial, because both axes involve the measured solid angle and galaxies 
span a limited range of dust temperatures.}
\label{fig:areavsbb}
\end{figure}

Total gas columns derived from interferometric CO or dust continuum 
observations \citep[e.g.,][]{sakamoto99}, fits to the IR to submm SED 
\citep[e.g.,][]{klaas01}, and modelling of OH absorptions 
\citep[e.g.,][]{gonzalezalfonso15} have been used to argue that
the dust emission in ULIRGs may be optically thick out to far-infrared 
wavelengths. Sizes and fluxes from \herschel\/-PACS can provide an independent
approach to this question. Fig.~\ref{fig:areavsbb} shows the ratio of the 
solid angle $\Omega_{BB}$ of a blackbody emitting half of the PACS flux 
at the temperature determined by the PACS photometry, and the 
measured half light solid angle at 70~$\mu$m $\Omega_{measured}$. 
$\Omega_{BB}$ is derived from the
Planck function for an optically thick blackbody reproducing both the
70 and 160~\mum\ (half) flux, and the measured solid angle is 
given by $\Omega_{measured}=\pi(R_{e,70}/D_A)^2$. 
This ratio will be low for optically thin 
emission or if small regions of higher optical depth are scattered over a 
larger area, but will approach unity for optically thick emission. In a 
strict sense
this is a lower limit, since emitting regions distributed in a 
way that is significantly different from the adopted Gaussian enlarge 
the measured solid angle for the same physical optical depth to the emission.
It also compares the measured size to a case that is optically thick over 
the full PACS wavelength range, and is hence a lower limit if the emission 
transits from optically thick to thin between 70 and 160~\mum\/. 

Several well known ULIRGs (Arp~220 which shows the largest value in 
Fig.~\ref{fig:areavsbb}, Mrk~231, Mrk~273, IRAS~F17207-0014, IRAS~13120-5453, 
see also Table~\ref{tab:compact})
show lower limits or measurements of 
$\Omega_{\rm BB}/\Omega_{\rm measured}\gtrsim$0.1, 
consistent with being optically thick in their central regions. The same
applies to some of the extremely compact circumnuclear regions in galaxies
with more modest \lfir\ (e.g., NGC~4418, IC~860, NGC~1266, among others).
In contrast, the majority of LIRGs and galaxies with lower \lfir\ is 
optically thin on average.

The case of Arp~220 may however also illustrate the complexities
of declaring a galaxy `optically thick in the far-infrared'. At 
$\Omega_{\rm BB}/\Omega_{\rm measured}=$0.35, Arp~220 is the most optically 
thick system with a
size measurement (rather than limit) in our sample, with a measured 
circularized \reblue\ of  246~pc (physical) or 0.67\arcsec\ (on sky). 
This is larger than the best evidence for the size of the central
star formation around the two nuclei: flickering supernovae have been
mapped in a series of VLBI papers \citep[][and references therein]{parra07}, 
and a similar area is suggested from 
free-free radio continuum by \citet{barcosmunoz15}, who give half light radii
of 51 and 35~pc around the two nuclei. A putative AGN would inject energy again
at very small scale. Both $\Omega_{\rm measured}$, which may still be inflated
by the presence of two semi-separated nuclei but a single Gaussian fit, and  
$\Omega_{\rm BB}$ are larger than what is suggested from the free-free 
continuum,
indicating the need for radiative transfer in the FIR from the energy sources
to the emitting surface. On the other hand, even shorter near-infrared
wavelengths that would be absorbed already by much lower dust columns 
provide a view of part of the stars around the two nuclei 
\citep[e.g.,][]{scoville00}, indicating a complex and patchy dust distribution.
A yet more extreme example is Mrk~231 with optical visibility towards its 
central type-1 AGN
despite $\Omega_{\rm BB}/\Omega_{\rm measured}>$0.11. These indicate a 
much more 
complex geometry with some rather transparent lines of sight, that are not 
expected in a spherically symmetric dust distribution around a 
heating source.

\subsubsection{A sample of compact far-infrared sources}

\begin{table*}
\caption{Compact far-infrared sources, selected by largest \sigfir\ and/or 
$\Omega_{\rm BB}/\Omega_{\rm measured}$}
\begin{tabular}{lcrrrrrrr}\hline
Name        &Morph& RA     & DEC    & z    &log(\lfir\/)&\reblue&log(\sigfir\/)&$\Omega_{\rm BB}/\Omega_{\rm m}$\\ 
                & &J2000   &J2000   &      &\lsun&kpc     &\lsun\/kpc$^{-2}$& \\
(1)&(2)&(3)&(4)&(5)&(6)&(7)&(8)&(9)\\\hline
III Zw 035      &D& 26.1280& 17.1021&0.0274&11.47&   0.272&   11.81&   0.05\\
NGC 1266        & & 49.0036& -2.4273&0.0072&10.32&$<$0.066&$>$11.89&$>$0.13\\
IRAS F05189-2524& & 80.2558&-25.3622&0.0426&11.75&   0.386&   11.78&   0.04\\
NGC 2623        & &129.6000& 25.7550&0.0185&11.41&   0.218&   11.93&   0.10\\
IRAS F09111-1007W&D&138.4016&-10.3245&0.0550&11.75&   0.456&   11.64&   0.10\\
IRAS F10173+0828& &155.0011&  8.2260&0.0491&11.57&   0.317&   11.77&   0.04\\
IRAS F12224-0624& &186.2663& -6.6811&0.0264&11.12&   0.175&   11.83&   0.11\\
NGC 4418        & &186.7277& -0.8776&0.0073&10.81&   0.056&   12.52&   0.14\\
Mkn 231         & &194.0590& 56.8738&0.0422&12.23&$<$0.386&$>$12.26&$>$0.11\\
IC 0860         & &198.7647& 24.6188&0.0112&10.82&   0.095&   12.07&   0.10\\
IRAS 13120-5453 & &198.7765&-55.1568&0.0308&12.10&   0.451&   11.99&   0.15\\
ESO 173-G015    & &201.8483&-57.4898&0.0097&11.42&   0.297&   11.67&   0.09\\
Mkn 273         & &206.1758& 55.8875&0.0378&11.93&$<$0.317&$>$12.13&$>$0.12\\
IRAS F14378-3651& &220.2455&-37.0754&0.0676&11.94&   0.465&   11.81&   0.05\\
CGCG 049-057    & &228.3046&  7.2257&0.0130&11.12&   0.163&   11.90&   0.15\\
Arp 220         & &233.7381& 23.5038&0.0181&12.11&   0.246&   12.53&   0.35\\
IRAS F17207-0014& &260.8416& -0.2838&0.0428&12.27&   0.564&   11.97&   0.10\\
NGC 7479        & &346.2359& 12.3229&0.0079&10.27&   0.118&   11.32&   0.23\\ \hline
\end{tabular}
\tablefoot{(2) 'D' denotes a pair/double galaxy (3),(4) Measured FIR position. 
For small and symmetric objects, this may be less accurate than the best 
literature position. (9) Ratio of solid 
angle of a blackbody emitting half of the FIR flux at the temperature fitting
the PACS photometry to measured half light solid angle.}
\label{tab:compact}
\end{table*}

Our data provide an opportunity to pick from a large sample of galaxies
an objective selection of the most compact far-infrared sources.
We do so by requiring either a large FIR surface brightness
\sigfir\/~$>10^{11.75}$~\lsun\/kpc$^{-2}$ or a large value of 
the optical depth
indicator $\Omega_{\rm BB}/\Omega_{\rm measured}>0.08$. The two criteria are
linked (inset to Fig.~\ref{fig:areavsbb}) and create overlapping samples, 
because dust temperatures only span a
limited range. Table~\ref{tab:compact} summarizes the resulting sample.
Unlike most of our analysis, we include here close galaxy pairs 
(marked `D' in the second column of Table~\ref{tab:compact}). If any, the 
emission from the offset second component in such a double would lower the
measured \sigsfr\ and 
optical depth. Presence of such a double in our sample of compact sources
hence implies that in the FIR they are dominated by one compact component.

While selected homogeneously and solely on the basis of FIR images, this sample
includes objects known to be peculiar and interesting in several ways.
Arp~220, IC~860, NGC~4418, for example have long been known to be 
`\cii\/-deficient' in comparison to their far-infrared luminosity 
\citep[e.g.,][]{malhotra01,luhman03}.  
Massive outflows of molecular gas have been detected 
in Mrk~231, Mrk~273, NGC~1266, IRAS F05189-2524,
IRAS~13120-5453, IRAS~F14378-3651, IRAS~F17207-0014 
\citep[][and references therein]
{fischer10,sturm11,veilleux13,gonzalezalfonso14,cicone14}. 
\sigfir\ of the sources in Table~\ref{tab:compact} corresponds
to \sigsfr$\gtrsim$100~\msun\/yr$^{-1}$kpc$^{-2}$ if due to star formation.
This is well above suggestions for the \sigsfr\ threshold for launching
powerful winds from local or high-z galaxy disks 
\citep{heckman02,newman12}. On the
other hand, AGN may play an important role in launching the most massive 
winds \citep{sturm11,veilleux13} and are present in several sources 
in Table~\ref{tab:compact}.
To which extent the conditions in and the phenomena related to these compact
regions relate to extreme density star-forming regions or embedded AGN is 
difficult to discern at many wavelengths, given the large average optical 
depth even in the FIR. Unless there are favourable unobscured lines of 
sight, the optical and near infrared emission will be limited to a surface
layer.  

Another well known galaxy with a compact FIR source is NGC~1377, which has 
been argued to host a nascent starburst \citep{roussel06}. We find a very 
small \reblue\/$<$60~pc far-infrared emitting region 
(Fig.~\ref{fig:bluesizeredshift}), but its modest log(\lfir\/)=9.7 and 
warm dust keep 
the lower limits to \sigfir\ and optical depth below the thresholds 
used to select Table~\ref{tab:compact}, despite the small size.

\subsection{The \cii\ deficit is closely linked to FIR surface brightness}
\label{sect:cii}

\begin{figure*}
\centering
\includegraphics[width=0.99\hsize]{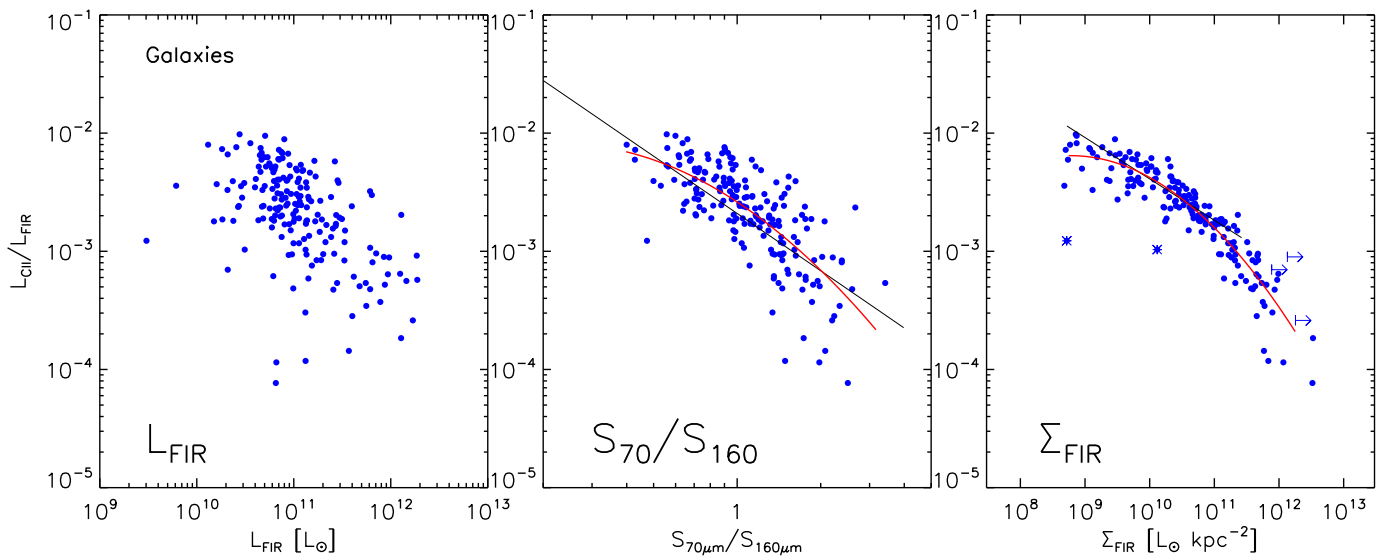}
\caption{Ratio of \cii\ and FIR luminosity as a function of 
FIR luminosity (left), FIR color (middle) and FIR surface brightness
(right panel). Red lines indicate the fits to our sample 
(Eq.~\ref{eq:ciifitcol} and \ref{eq:ciifit}), and black lines
results of \citet{diazsantos13}.
The two slightly outlying points marked with an overlayed
asterisk in the right panel were not used for the fit that is shown in red.
Suitable \cii\ data are available for the 182 galaxies 
that are plotted here.}
\label{fig:cii}
\end{figure*}

The \cii 158~\mum\ line typically is the major coolant of neutral 
interstellar gas. It is detectable with current instrumentation to 
redshifts above 6 and has been proposed as an SFR tracer 
\citep[e.g.,][]{stacey91,delooze11,herreracamus15}. Several known effects
add uncertainty to the use of \cii\ for that purpose. First, low metallicity 
galaxies can show enhanced \cii\ from regions where \cii\ coexists with
self-shielded H$_2$ while the smaller amount of dust implies that carbon 
is in the form of CO only in a smaller region 
\citep[e.g.,][]{madden97}. We do not
address this effect further in this work. \cii\ can 
also be bright in the atomic and ionized interstellar media of the outer 
parts of galaxies \citep{madden93}, with a less tight link to
ongoing star formation than for PDRs around regions of active star formation. 
Finally, progressing from normal
galaxies to LIRGs and ULIRGs a reduced ratio \cii\//FIR for luminous 
objects has been reported -- the `\cii\ deficit' 
\citep[e.g.,][]{malhotra97,malhotra01,luhman98,luhman03,fischer14}. 
The deficit can also be linked
to offset from the main-sequence \citep{graciacarpio11,diazsantos13}, a
view that reconciles the \cii\ properties of high and low redshift galaxies
that differ if compared at same IR luminosities. Physical explanations for the
deficit discussed in these references include (i) a reduced photoelectric 
heating efficiency at high ratios of UV radiation field and gas density, 
(ii) related to this, dust competing for photons in high ionization 
parameter `dust-bounded' H{\sc ii} regions, (iii) AGN contributions to the 
FIR continuum, and (iv) self-absorbed \cii\ emission.

We have supplemented our galaxy sample with \cii\ fluxes from the literature.
Most are based on the measurements of \citet{diazsantos13}. We have
scaled their \cii\ fluxes which are based on the central spaxel of the 
PACS integral field spectrometer by the ratio of our total 160~\mum\ flux 
density and their 157~\mum\ continuum flux density, dropping
objects with correction factor above 10 or 160~\mum\ size above 40\arcsec\/, 
where this correction is too uncertain. We also use \cii\ fluxes from 
\citet{brauher08}, excluding galaxies that were flagged as extended in that 
paper, and those with a measured 160~\mum\ size above
the ISO-LWS beam size of 80\arcsec\/. Finally, a few objects are added from
\citet{sargsyan12} and \citet{farrah13}.

Fig.~\ref{fig:cii} shows the ratio of \cii\ and FIR emission as a function
of three different quantities. The left panel shows for our sample the 
classical `deficit' at high FIR luminosities. The scatter is large, however,
and some of the most extreme deficits arise at intermediate FIR luminosities, 
as found previously \citep[e.g.,][]{malhotra97,malhotra01}. The middle panel 
shows a somewhat improved
trend as a function of far-infrared color, again confirming the earlier work
referenced above. Our data can be fitted by
\begin{equation}
\begin{split}
log(L_{\rm [CII]}/L_{\rm FIR}) & = & -2.583 - 1.551\times log(S_{70}/S_{160})\\ 
                              & & - 1.220\times (log(S_{70}/S_{160}))^2,\\
\end{split}
\label{eq:ciifitcol}
\end{equation}
with the fit applicable to the range $log(S_{70}/S_{160})\sim -0.4\ldots 0.5$.
The dispersion around this fit  is 0.27~dex, virtually identical to
the 0.275~dex that we obtain for our sample around the relation 
(for a color $S_{63}/S_{157}$) in Eq.~1 of \citet[][]{diazsantos13}, 
or the 0.28~dex that 
these authors quote for their sample which has significant overlap with ours,
but studies lines and continuum from the central spaxel of the PACS 
integral field spectrometer.

\begin{figure*}
\centering
\includegraphics[width=0.99\hsize]{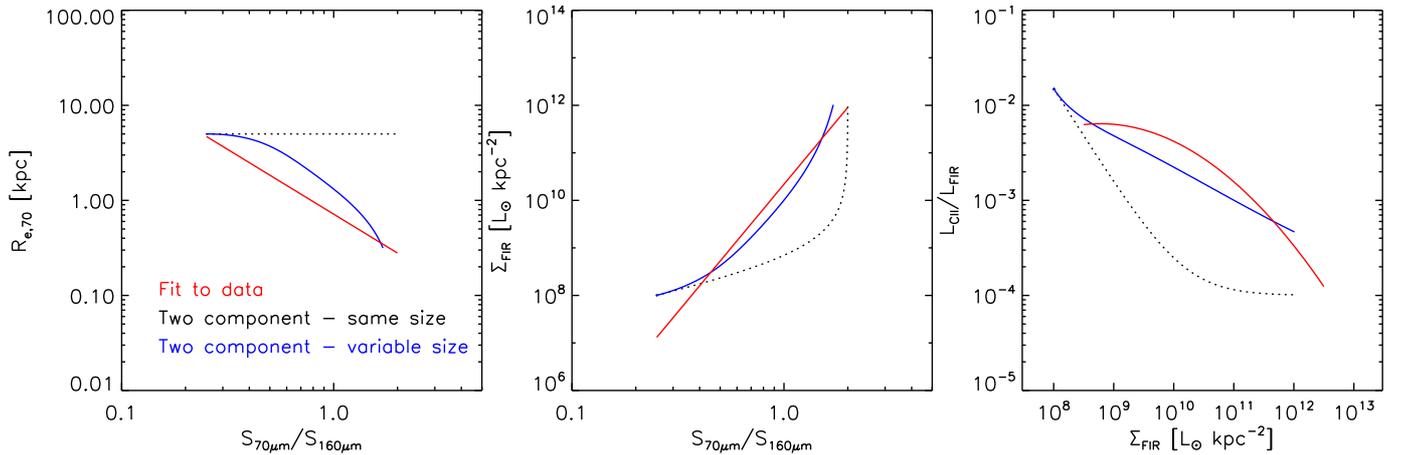}
\caption{Comparison of observed scalings to two component toy models.
Red lines show the fits to the data as previously shown in 
Fig.~\ref{fig:colorscalings} (here: left and center panel) and 
Fig.~\ref{fig:cii} (right panel). Both toy models include a mix of a `disk' 
component with cold dust and a high ratio \cii\//FIR, and a `dense star 
formation' component with warm dust and low  \cii\//FIR, with varying surface 
brightness of the warm component. Spreading the dense SF component with low 
filling factor over a large disk is a very poor fit, while scaling down its 
size with surface brightness gives a first approximation to the data.}
\label{fig:mix}
\end{figure*}

An extremely tight relation is 
observed between the ratio of \cii\ and FIR emission and FIR surface brightness
(right panel). The red line shows a simple quadratic fit to the measurements
and limits, where we have not 
used for the fit the two slightly outlying sources 
NGC~5866 and MCG-03-34-064 (marked with asterisk extensions to the
symbols in the right panel of Fig~\ref{fig:cii}). We cannot determine with the
data used here whether these are true outliers, or whether they are affected 
by technical issues, e.g., an unusually extended \cii\ flux distribution.
The relation shown in the right panel of Fig.~\ref{fig:cii} is
\begin{equation}
\begin{split}
log(L_{\rm [CII]}/L_{\rm FIR}) & = & -11.7044 + 2.1676\times log(\Sigma_{\rm FIR})\\ 
                              & & - 0.1235\times (log(\Sigma_{\rm FIR}))^2.\\
\end{split}
\label{eq:ciifit}
\end{equation}
The dispersion around this relation, covering 3.5~dex in surface brightness 
and 2~dex in the ratio of \cii\ and FIR, is 
0.16~dex for the full sample and 0.15~dex if excluding the two outliers.
\citet{diazsantos13,diazsantos14} used sizes measured in the mid-infrared to 
derive a log-linear relation of deficit and IR surface
brightness \citep[Eq. 4 of][]{diazsantos13} which we also overplot to 
Fig~\ref{fig:cii} right, corrected for an adopted \lir$=$1.9$\times$\lfir\/. 
This relation was derived for a clearly star formation dominated (U)LIRG
subsample, thus minimizing a potential distortion of mid-IR based sizes
by AGN emission. Because of the more uncertain size 
measures and more size limits, this relation covers a smaller range 
\sigfir$\approx 5\times 10^{8}\ldots 3\times 10^{11}$\lsun kpc$^{-2}$.   
Over this restricted range, the two relations agree well, and have small 
dispersions: 0.15~dex quoted by \citet{diazsantos13}, 0.14~dex for our data 
(minus the two outliers) around their relation, and 0.12~dex for our data and
our relation. Over the wider range in surface brightness that is accessible
via the \herschel\ size measurements, Fig~\ref{fig:cii} shows the 2nd order
fit to be more appropriate than the log-linear relation. This is also
reflected in the dispersions which are 0.19~dex around the log-linear relation
vs 0.15~dex around the fit of Eq.~\ref{eq:ciifit}.    
 
Existence of such a relation is clearly expected from earlier work on the
\cii\ deficit in combination with the scalings presented in this paper.
From resolved observations of nearby 
galaxies, \citet{herreracamus15} report a scaling 
$\Sigma_{\rm SFR}\propto \Sigma_{\rm [CII]}^{1.13}$ for a regime of low star 
formation surface density $\Sigma_{\rm SFR}\lesssim 0.1$~\msun yr$^{-1}$. This 
suggests a relatively flat continuation towards lower surface FIR brightness 
of the relation shown in the right panel of Fig.~\ref{fig:cii}, 
for a range that is not 
covered by the combination of our sample with the specific 
\cii\ literature data.

The noteworthy aspect of this relation is its remarkable tightness. Given that
our far-infrared surface brightness is a galaxy average, one might have 
expected a tighter correlation of the deficit with FIR color, which is 
representing the local physical conditions in and around star-forming 
complexes, but this is not the case (Fig.~\ref{fig:cii}). A large number of 
studies have argued that even in the ULIRGs of our sample, the total infrared
(and even more the far-infrared) luminosity is typically not dominated by an 
AGN, even where present \citep[e.g.,][]{genzel98,lutz98,veilleux09a}. 
$\Sigma_{\rm FIR}$ can then be directly linked to $\Sigma_{\rm SFR}$,
and is not strongly affected by the AGN. 
The tightness
of the link between \cii\ deficit and \sigfir\ then  makes a dominant role 
of scenario (iii) above -- simple dilution by AGN-heated FIR emission -- 
unlikely. This
is in line with the conclusion from comparisons to various AGN indicators 
\citep{diazsantos13}. 

We are thus directed at explanations that are related to the physical
conditions in star-forming regions as well as the diffuse ISM. A
simple test that could represent the varying conditions is to construct a 
two component mixed model with 
(1) strong \cii\ and cold dust, representative of a more diffuse galaxy disk, and 
(2) dense star formation with weak \cii\ and warm dust, to fit 
the observed trends. Such simple scenarios have often been invoked, starting
from the `cirrus plus star formation' interpretation of IRAS color-color 
diagrams. For a toy model, we assign to the cold component 
S$_{70}$/S$_{160}$=0.25, \cii\//FIR=0.015, \sigfir =10$^{8}$~\lsun kpc$^{-2}$
and a Gaussian with size \reblue =5~kpc.
For the dense SF component, we adopt S$_{70}$/S$_{160}$=2 and 
\cii\//FIR=0.0001. These values are picked on the basis of 
Fig.~\ref{fig:colorscalings} and Fig.~\ref{fig:cii} as well as the \cii\ 
literature. We then vary the surface brightness of the dense SF component from
zero to a maximum of \sigfir =10$^{12}$~\lsun kpc$^{-2}$. A first toy model
is to simply add the dense SF component over the entire disk, as if more
and more additional small but intense star-forming regions were spread 
over the disk.
This is a very poor representation of the data (black dotted lines in 
Fig.~\ref{fig:mix}). Not 
only are size trends not captured by definition, but the fits to the trend of 
color and  \cii\//FIR with surface brightness are very poor, because the 
dense SF component quickly dominates as its surface brightness increases.
In a second model we assume that the `few dense SF regions scattered 
over the entire disk' scenario only applies for low surface brightness of that
component, but that high surface brightness of the dense
SF can only be reached if the limited gas content of local galaxies is 
compressed into a small region. We implement this by keeping the
disk component fixed as a Gaussian with \reblue =5~kpc, and add to it
a second Gaussian for the dense SF component with size scaling as
$R_{\rm e,70} = 5 \times (1+\Sigma_{\rm FIR}/10^8)^{-0.3}$ with \reblue\ in 
kpc and \sigfir\ in \lsun kpc$^{-2}$. This parametrization gives the
dense SF the same size as the cold disk as long as its surface brightness is
low, but shrinks it at high surface brightness, as motivated above. The 
value -0.3 for the power law slope 
is based on a manual adjustment, aiming at a reasonable fit to the 
Fig.~\ref{fig:mix} trends. Size and surface brightness are then 
derived by single Gaussian fitting this mix, as for the real data. This
second toy model gives a more reasonable approximation of the observed
trends (blue lines in Fig~\ref{fig:mix}). Given its simplicity, we refrain 
from further modifying the assumptions towards a better fit.

The relative success of the toy model that scales the size of the dense SF
component should not be overinterpreted. It uses simplified observational 
facts and no ISM physics. But it provides some confidence that a coherent
interpretation of the trends discussed in this paper should be possible
via models implementing more realistic distributions of ISM and star 
formation, as well as the physics of the dusty and gaseous phases of the
H{\sc ii} regions, PDRs, and diffuse ISM. Likely, such models will invoke
more smooth trends of properties than derived from coadding two extremes in 
the toy model, and a \cii\//FIR that only decreases at somewhat higher 
surface brightness. The goal would be to use physical models to quantitatively 
relate the trends of color and \cii\//FIR to the increase in typical 
radiation field intensity at higher FIR surface brightness and smaller size, 
while using plausible assumptions on ISM structure.

\begin{figure*}
\centering
\includegraphics[width=0.49\hsize]{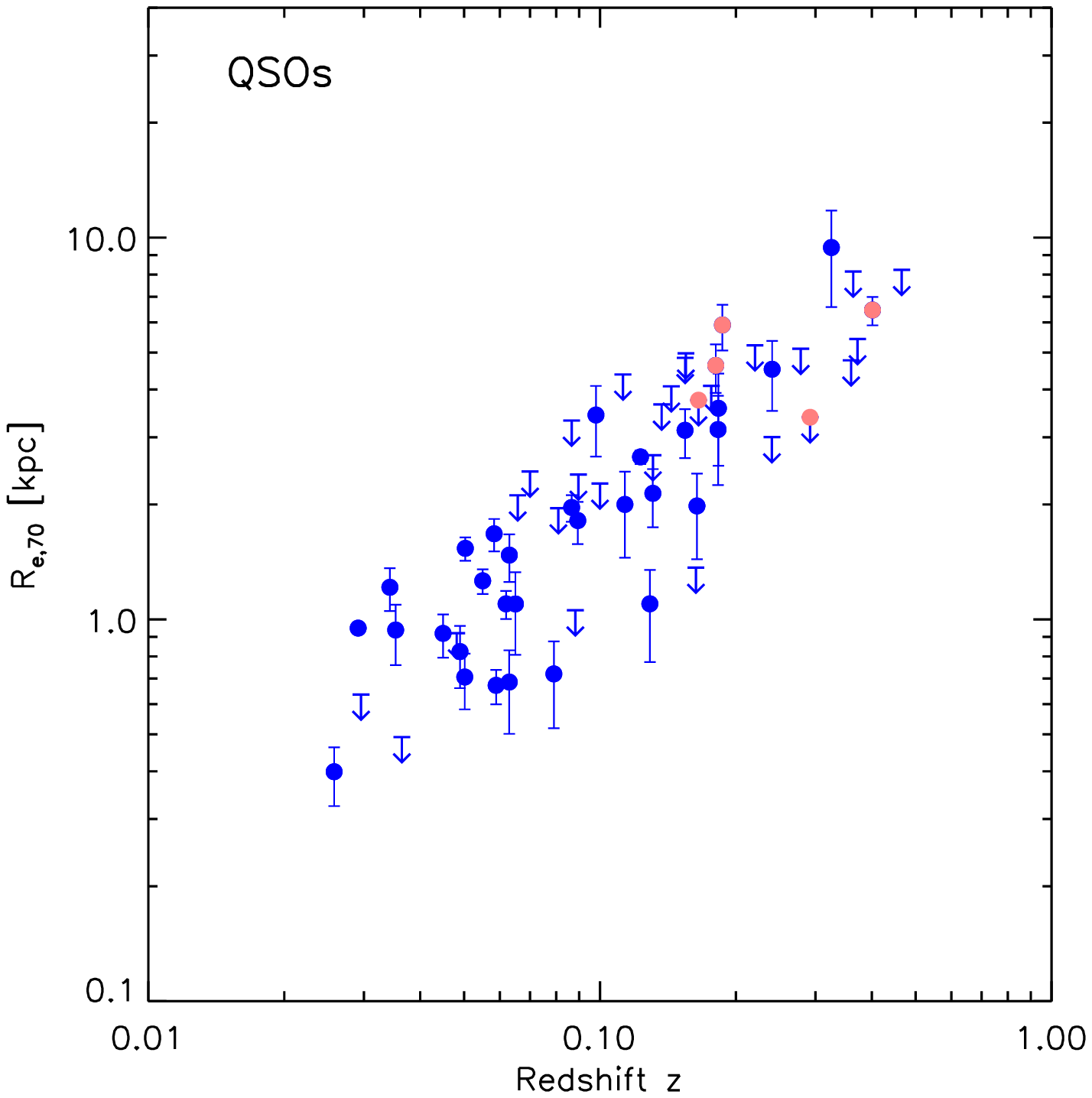}
\includegraphics[width=0.49\hsize]{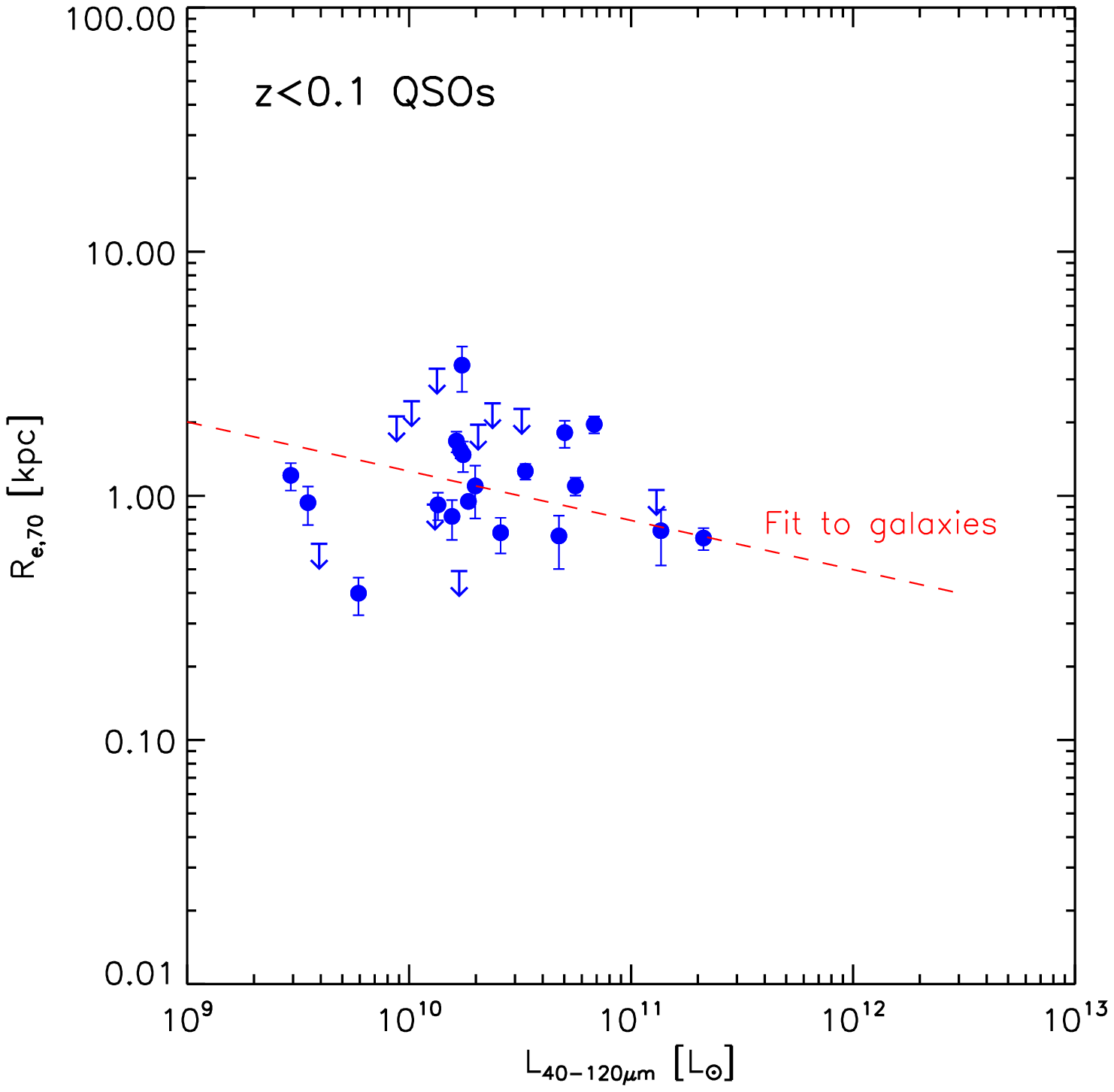}
\caption{Left: Half light radius at 70~\mum\ as a function of redshift 
for PG QSOs 
and their hosts. Pink symbols mark systems with literature evidence for
being double/interacting, which may inflate the measured size. Right:
Half light radius at 70~\mum\ vs. FIR luminosity for the z$<$0.1 subsample 
of QSOs
with favourable ratio of size measurments to limits. The relation
derived above for {\emph galaxies} is overplotted.
}
\label{fig:pgsize}
\end{figure*}

\begin{table}
\caption{Summary of scalings}
\begin{tabular}{lc}\hline
Scaling             & Section\\ \hline 
\\                   
Scalings with \lfir :& \\
$log (R_{\rm e,70}) = 0.101 - 0.202\times (log(L_{\rm FIR})-10)$&\ref{sect:scalings}\\
$log (\Sigma_{\rm FIR})  = 8.997 + 1.408\times (log(L_{\rm FIR})-10)$&\ref{sect:scalings}\\
$log (\Sigma_{\rm SFR})  =  -1.117 + 1.408\times log({\rm SFR})$&\ref{sect:scalings}\\
\\
Scalings with main-sequence offset:&\\
$log (R_{\rm e,70}) = 0.19 - 0.39\times log({\rm sSFR}/{\rm sSFR}_{\rm MS})$&\ref{sect:scalings}\\
$log (\Sigma_{\rm FIR}) = 8.19 + 2.59\times log({\rm sSFR}/{\rm sSFR}_{\rm MS})$&\ref{sect:scalings}\\
$log (\Sigma_{\rm SFR}) = -1.53 + 2.59\times log({\rm sSFR}/{\rm sSFR}_{\rm MS})$&\ref{sect:scalings}\\
\\
\multicolumn{2}{l}{Scalings with far-infrared color and dust temperature:}\\
$log (R_{\rm e,70}) = -0.145 - 1.362\times log(S_{70}/S_{160})$&\ref{sect:scalings}\\
$log (\Sigma_{\rm FIR}) = 10.34 + 5.37\times log(S_{70}/S_{160})$&\ref{sect:scalings}\\
$log (\Sigma_{\rm SFR}) = 0.62 + 5.37\times log(S_{70}/S_{160})$&\ref{sect:scalings}\\
$log (\Sigma_{\rm FIR}) = -22.5 + 21.8\times log(T_{\rm d})$&\ref{sect:scalings}\\
\\
\multicolumn{2}{l}{Scalings of \cii\//\lfir\ with far-infrared color and far-infrared}\\
\multicolumn{2}{l}{surface brightness:}\\
$log(L_{\rm [CII]}/L_{\rm FIR}) = -2.583 - 1.551\times log(S_{70}/S_{160})$&\\ 
\hspace{2.5cm}$\hbox{} - 1.220\times (log(S_{70}/S_{160}))^2$&\ref{sect:cii}\\
$log(L_{\rm [CII]}/L_{\rm FIR})  =  -11.7044 + 2.1676\times log(\Sigma_{\rm FIR})$&\\  
\hspace{2.5cm}$\hbox{} - 0.1235\times (log(\Sigma_{\rm FIR}))^2$&\ref{sect:cii}\\ 
\\ \hline
\end{tabular}
\tablefoot{See the respective section for details. Scalings for \sigsfr\
are a simple linear conversion of the observed \sigfir\ scalings.}
\label{tab:scalings}
\end{table}

\subsection{A similar size of the FIR emission in QSO hosts and galaxies}
\label{sect:qsosize}

Because of the favourable contrast between the SED of an AGN and the SED 
of a star-forming galaxy \citep[e.g.,][]{netzer07}, rest frame far-infrared 
emission has been widely used as a 
star formation tracer in AGN hosts, in particular in the context of \herschel\
surveys \citep{lutz14} up to z$\sim$2 \citep[e.g.,][]{rosario12}. 
While it is often safe to 
simply ascribe this FIR emission to star formation, this is no longer the 
case for powerful AGN in hosts with low SFR. Both \herschel\ color arguments
\citep{hatziminaoglou10,rosario12} and attempts to construct local `intrinsic'
AGN SEDs reaching out to the FIR \citep{netzer07,mullaney11,mor12} have
been used to delineate the border between these two regimes. Comparing the 
size of the FIR emission in QSO hosts with that in other galaxies can provide 
another constraint to this problem, in addition to information on the
host and/or AGN proper.

Fig.~\ref{fig:pgsize} left shows results for 59 QSOs (32 size measurements
and 27 size limits). The QSOs are on average much more distant than the 
galaxies discussed before, leading to a less favourable ratio of \reblue\ 
detections to limits. Also, our approach of attempting size measurements only
for photometric S/N$>$10 has already shrunk the sample from 93 QSOs observed
by \herschel\ to these 59, which will miss some of the FIR-weak and distant
objects. Most \reblue\ measurements span a range $\sim 0.5\ldots 5$~kpc with a 
median of 1.7~kpc (1.1~kpc for the z$<$0.1 subset discussed below). Limits are 
consistent with this range. Exceptionally, we plot here also (marked in pink)
five QSOs where NED and literature give a warning of a close double
nature that could inflate the size if both components emit in the FIR.
Only in PG1543+489 does the \herschel\ image provide direct evidence for that,
but we discard all five from further analysis.

In Fig.~\ref{fig:pgsize} left, there appears to be a lack of 
distant but small QSOs and of local but large QSOs. The first 
category, to the extent it may be present in the sample, would simply be 
assigned upper size limits that increase with redshift.
The absence of large \reblue $\lesssim 10$~kpc local hosts is more 
noteworthy and related
to the known heterogeneity of the QSO host population, on which 
optical/near-infrared studies provide some insights. The hosts of
z$<$0.06 low luminosity QSOs show a mix of disks, ellipticals, and merger 
remnants, with typical near-infrared $R_e \sim 2$~kpc and none with  
$R_e > 5$~kpc \citep{busch14}. In contrast, the hosts of the luminous 
z$\sim$0.2 QSOs
studied by \citet{dunlop03} are typically  $R_e \sim 10$~kpc giant ellipticals.
The PG sample, approximately UV/optical flux limited and covering a range
of redshifts, bridges these two regimes -- it includes a few large hosts,
but only in the large volume accessible at z$>$0.1 \citep{veilleux09, 
dunlop03}. On the other hand, many PG QSOs may be on an evolutionary path
from IR luminous mergers to moderate size `disky' ellipticals 
\citep[e.g.,][]{dasyra07}.
Already in the optical/NIR tracers of the stellar population, the z$<$0.1
region lacks the rare $R_e \sim10$~kpc QSOs, in plausible agreement
with the lack of large FIR sizes in our sample, and the ratios of FIR to NIR
size $\lesssim$1 that we report below.

To better compare with other galaxies, we restrict the QSO sample to the
33 z$<$0.1 objects. Of those, only 4 were not fit because of S/N<10,
and the number of size measurements among the remaining 29 is high: 19 vs.
10 upper limits. 
Compared to the full PG sample, this redshift limit and the photometric S/N 
cut prefer moderate AGN luminosities and exclude the IR faintest objects, 
specifically the median log($L_{\rm Bol,AGN}$) is 44.8 and the median 
log(\lfir\/) is 10.2. 
Fig.~\ref{fig:pgsize} right shows their \reblue\ vs. \lfir\/, and overplotted
the fit relation derived above for {\em galaxies} (Eq.~\ref{eq:size_fir}). 
Clearly, the \reblue\ 
sizes for the QSOs are fully consistent with those of other local galaxies of 
the same infrared luminosity. This is consistent with their FIR emission being
due to star formation in the host, giving support to the use of FIR emission
as a SFR indicator in this regime of \lfir\//$L_{\rm Bol,AGN}\approx$0.1.
For QSOs with a lower ratio of far-infrared and bolometric luminosity that
are not sampled by our size measurements, AGN heated and/or `cirrus' dust 
may be relatively more important, and a similar size consistency should not
be taken for granted.

\begin{figure}
\centering
\includegraphics[width=\hsize]{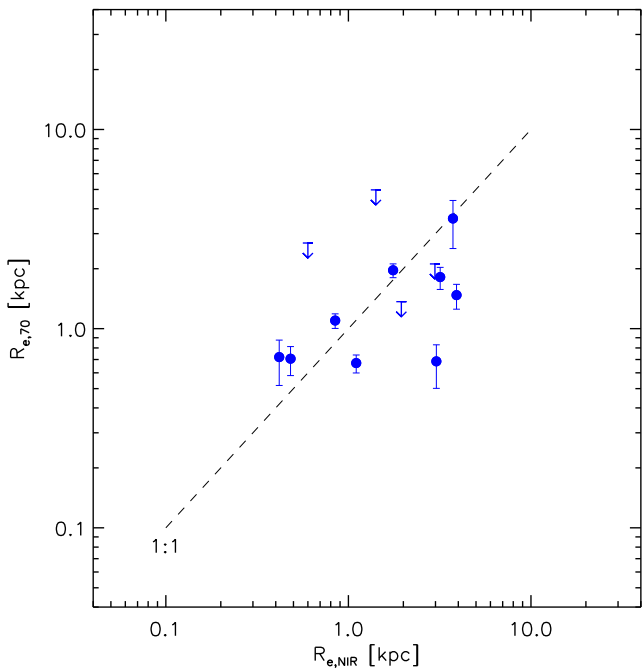}
\caption{Comparison of half light radii measured in the far-infrared 
(this work) with near-infrared half light radii for the same sources
(Table~5 of \citet{veilleux09}). The dashed line visualizes a 1:1 relation,
it is not a fit.
}
\label{fig:pgnirfirsize}
\end{figure}

\begin{figure}
\centering
\includegraphics[width=\hsize]{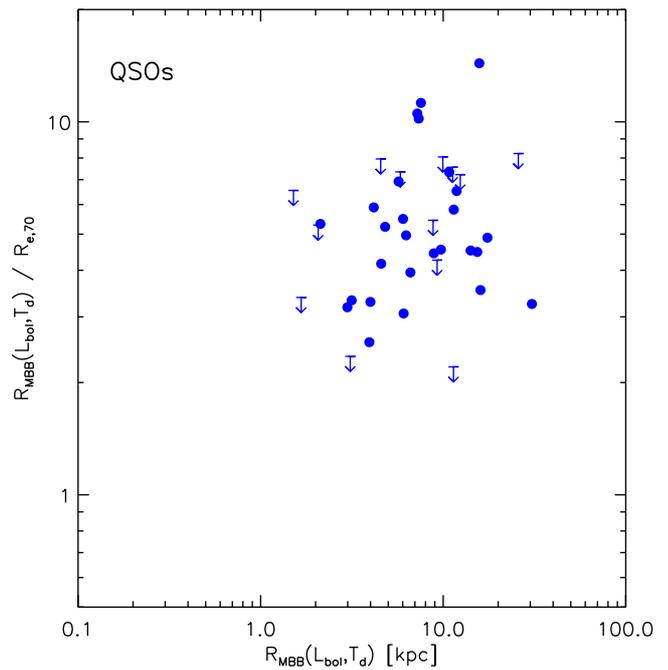}
\caption{Comparison of the radius at which an AGN of the given QSO's luminosity
heats directly exposed optically thin dust to the measured dust temperature,
and the measured far-infrared half light radius.
}
\label{fig:pgradlbol}
\end{figure}

We may also compare the size of the far-infrared emission with the size of 
the stellar host as seen in the near-infrared. \citet{veilleux09} present half
light radii of the near-infrared emission for a sample of PG QSOs,
as observed with HST-NICMOS, and after subtracting the AGN point source.
Fig.~\ref{fig:pgnirfirsize} compares NIR and FIR half light radii for
13 QSOs in that sample with \herschel\ size measurement 
(9/4 measurements/limits). The comparison suggests FIR size similar to or 
in some cases smaller than the stellar component, again consistent with 
FIR emission originating in the host, and star formation taking place on a 
spatial scale similar to the pre-existing stellar population.

Finally, we can compare the measured \reblue\ to the radius at which the AGN
can heat dust to the measured temperature. At the smaller $\ll$kpc scales 
of the circumnuclear obscuring structure discussed in the context of AGN 
unification, dust will be warm and the emission mostly
in the mid-infrared. This is what is empirically found to dominate 
AGN-heated dust SEDs \citep[e.g.,][]{netzer07}. But 
strong far-infrared emission could emerge if a 
substantial amount of dust is exposed to the AGN radiation only at large 
distances, without the dominant UV component being intercepted by a modest 
amount of dust further in. This could occur for example in a strongly warped
disk  \citep{sanders89}.
Fig.~\ref{fig:pgradlbol} compares the measured \reblue\ sizes to the radius
at which the AGN heats optically thin dust to the measured dust temperature.
Here we have adopted AGN bolometric luminosities based on the 5100~\AA\ 
continuum of the PG QSOs and the bolometric correction of Eq.~7.3 in 
\citet{netzer13}.
For the dust properties in Eq. 5.85 of \citet{netzer13} the expected 
radius in kpc is
$R_{\rm MBB}=2.35\times 10^{5} L_{46}^{0.5} T_{\rm MBB}^{-2.6}$. Here, $L_{46}$ is
the bolometric luminosity of the AGN in units of 10$^{46}$\ergs\ and $T_{MBB}$
is derived by fitting a modified blackbody to the PACS photometry, adopting an emissivity index $\beta = 1.2$ for consistency with the dust properties
adopted by \citet{netzer13}. 
Fig.~\ref{fig:pgradlbol} shows that the radii estimated this way 
are typically a factor $\sim$5 larger than the measured \reblue\/, and also
uncomfortably large compared to typical NIR host sizes as seen in 
Fig.~\ref{fig:pgnirfirsize}. Dust at large radii that is directly heated by 
AGN light that has not been absorbed further in
seems hence an unsatisfactory explanation for the measured FIR sizes. It is
beyond the scope of this work to study to which extent detailed radiative 
transfer of the incident direct AGN UV and/or infrared reemission 
can avoid this problem with 
FIR sizes, without violating SED constraints. At this point, the analogy 
of QSO hosts to 
galaxies (Fig.~\ref{fig:pgsize}) seems the more natural explanation.
The FIR emitting dust will here be heated mostly host star formation
on scales similar to other galaxies, and in regions where the radiation 
field is not dominated by the AGN, for example outside
the ionisation cone. In this scenario, the direct AGN radiation would be 
intercepted already at smaller
radii by the obscuring material invoked by unified AGN models, and re-emitted 
mostly in the mid-IR.

Among the PG QSOs with largest \reblue\ in Fig.~\ref{fig:pgsize} left,
several are marked as double. This implies that the size measurement
might be inflated by a companion. For PG1543+489 
we see direct evidence in the form of a weak southern extension 
in the 70~\mum\ image, consistent with a companion that is present in archival 
NICMOS images. 
No literature evidence for a strong companion exists for the largest QSO
source,
PG2251+1113 with \reblue$\sim$9.4kpc, similar to the largest non-QSO
galaxies in our sample. This is a quite powerful QSO 
($L_{\rm Bol,AGN}\sim 10^{46.5}$\ergs\/) which has made studies of
the stellar component in the near-infrared difficult. The fits of
\citet{guyon06} and \citet{veilleux09} suggest $R_{\rm e,NIR}\gtrsim 10$kpc.
Due to the difficulty of subtracting the point source they are considered 
unreliable, but provide some support to the very large FIR size. 

\citet{mushotzky14} present initial results from a \herschel\ study, 
including PACS far-infrared size information, for the BAT AGN sample that 
is selected in
ultra-hard X-rays. The BAT sample covers AGN luminosities and redshifts
typically lower than the PG QSOs, but reaches up to the level of the PG QSOs.
Measured far-infrared sizes span a wide range, with about a third
of size limits \reblue $\lesssim$1~kpc but also some fairly extended objects
with IR surface brightness lower than KINGFISH galaxies. For resolved sources,
160~$\mu$m sizes tend to exceed the 70~$\mu$m sizes. This is in broad agreement
with our findings for galaxies and QSOs. A systematic comparison of
this sample covering a wide range of AGN luminosities to \lfir\/-matched 
non-AGN galaxies will be worthwhile to extend the limited range covered by 
PG QSOs (Fig.~\ref{fig:pgsize}).

In summary, the FIR size of the PG QSOs for which we
could attempt measurements seems to resemble that of galaxies of same
FIR luminosity, consistent with being dominated by star formation in the host.
It is important to recall that this statement refers to a subset of modest 
luminosity PG QSOs: log($L_{\rm Bol,AGN}$)$\approx$44.8 and 
\lfir\//$L_{\rm Bol,AGN}$$\approx$0.1. It should not be blindly transferred to
systems with smaller \lfir\//$L_{\rm Bol,AGN}$.

\section{Conclusions}

We have used \herschel\ 70, 100, and 160~\mum\ images to study the size of the
far-infrared emission in a sample of 399 local galaxies and QSOs. We rely on
the stable point spread function and subtraction in quadrature to infer half 
light radii of the sources, reaching well below the PSF half 
width for sources with good S/N. We find that:
\begin{itemize}
\item Galaxies with \lfir$\sim$10$^{11}$\lsun\ can be found with very different 
distributions of infrared emission (hence star formation). These range from
large R$_{\rm e}\sim$10~kpc disks down to compact $<$100~pc circumnuclear 
bursts.
In contrast,  \lfir$\gtrsim 10^{12}$~\lsun\ ULIRGs are only found with compact
R$_{\rm e}\sim 0.5$~kpc morphologies, likely due to the need to compress the
limited gas content to temporarily achieve the large IR luminosities.
On average, but with large scatter, far infrared half light radius
scales with \lfir$^{-0.2}$ and far-infrared surface brightness (SF surface 
density) with  
\lfir$^{1.4}$. First \alma\ measurements suggest a larger size of high-z 
sources at equivalent \lfir\/, albeit at slightly longer rest wavelengths. 
\item Half light radius and FIR surface density also show clear trends 
when moving from the main-sequence of star formation to higher sSFR, 
with logarithmic slopes of -0.4 and 2.6, respectively.
\item There is a fairly tight relation of both half light radius and FIR
surface brightness with FIR color (dust temperature). There is consistency 
when comparing the scalings of FIR surface brightness with dust temperature 
with expectations from the scaling of \sigfir\ with main-sequence 
offset, combined with the \citet{genzel15} scaling of gas depletion time with 
main-sequence offset and the assumption that dust is an optically thin 
calorimeter.        
\item The average optical depth over the size of the far-infrared emitting
region is large in some LIRGs and ULIRGs.
\item The ratio of \cii\ 158~\mum\ and FIR emission and the `\cii\ deficit' is 
more tightly linked to FIR surface brightness than to FIR luminosity or 
FIR color. For our sample, the dispersion is about 0.15~dex for a relation 
covering 3.5 dex in surface brightness. 
\item The size of the far-infrared emission in z<0.1 PG QSOs with 
\lfir\//$L_{\rm Bol,AGN}\approx$0.1 is consistent 
with that of galaxies with same FIR luminosity. This is consistent with 
host star formation creating their far-infrared emission.    
\end{itemize}

\begin{acknowledgements}

We thank the anonymous referee for a careful review and helpful comments. This 
work would not have been possible without the efforts at the PACS ICC and 
Herschel science center that pushed the accuracy of \herschel\ pointing
reconstruction beyond its limits. In particular, we would like to thank 
Herv\'e Aussel, Helmut Feuchtgruber, Craig Stephenson, and Bart Vandenbussche.
HN acknowledges support by ISF grant no. 284/13.
PACS has been developed by a consortium of institutes led by MPE
(Germany) and including UVIE (Austria); KUL, CSL, IMEC (Belgium); CEA,
OAMP (France); MPIA (Germany); IFSI, OAP/OAT, OAA/CAISMI, LENS, SISSA
(Italy); IAC (Spain). This development has been supported by the funding
agencies BMVIT (Austria), ESA-PRODEX (Belgium), CEA/CNES (France),
DLR (Germany), ASI (Italy), and CICYT/MCYT (Spain).
\end{acknowledgements}

\setlength{\tabcolsep}{0.06cm}
\begin{longtab}
\begin{longtable}{lrrccrrrrrrrrr}
\caption{\label{tab:longtable}Far-infrared size and surface brightness of galaxies and QSOs} \tabularnewline\hline
Name&RA&DEC&Cl&Mo&Scale&\multicolumn{2}{c}{\reblue}&\multicolumn{2}{c}{\regreen}&\multicolumn{2}{c}{\rered}&\multicolumn{2}{c}{log(\sigfir)}\tabularnewline
&J2000&J2000&&&kpc/\arcsec&\multicolumn{2}{c}{kpc}&\multicolumn{2}{c}{kpc}&\multicolumn{2}{c}{kpc}&\multicolumn{2}{c}{\lsun kpc$^{-2}$}\tabularnewline
(1)&(2)&(3)&(4)&(5)&(6)&\multicolumn{2}{c}{(7)}&\multicolumn{2}{c}{(8)}&\multicolumn{2}{c}{(9)}&\multicolumn{2}{c}{(10)}\tabularnewline\hline
\endfirsthead
\hline
Name&RA&DEC&Cl&Mo&Scale&\multicolumn{2}{c}{\reblue}&\multicolumn{2}{c}{\regreen}&\multicolumn{2}{c}{\rered}&\multicolumn{2}{c}{log(\sigfir)}\tabularnewline 
&J2000&J2000&&&kpc/\arcsec&\multicolumn{2}{c}{kpc}&\multicolumn{2}{c}{kpc}&\multicolumn{2}{c}{kpc}&\multicolumn{2}{c}{\lsun kpc$^{-2}$}\tabularnewline
(1)&(2)&(3)&(4)&(5)&(6)&\multicolumn{2}{c}{(7)}&\multicolumn{2}{c}{(8)}&\multicolumn{2}{c}{(9)}&\multicolumn{2}{c}{(10)}\tabularnewline\hline
\endhead
\hline
\endfoot

UGC 12914           &   0.4110&  23.4825& G&  & 0.297&   5.876&$\pm$0.032&   5.989&$\pm$0.018&   6.348&$\pm$0.020&   7.82&$\pm$0.00\tabularnewline
UGC 12915           &   0.4256&  23.4956& G&  & 0.295&   2.086&$\pm$0.003&   2.200&$\pm$0.003&   2.278&$\pm$0.014&   9.06&$\pm$0.00\tabularnewline
PG 0003+158         &   1.4968&  16.1634& Q&  & 5.766&        &          &        &          &        &          &        &        \tabularnewline
PG 0003+199         &   1.5805&  20.2027& Q&  & 0.519&   0.399&$\pm$0.075&   0.693&$\pm$0.105&   1.556&$\pm$0.314&   9.77&$\pm$0.12\tabularnewline
NGC 23              &   2.4716&  25.9237& G&  & 0.310&   1.004&$\pm$0.003&   1.125&$\pm$0.006&   1.567&$\pm$0.023&   9.94&$\pm$0.00\tabularnewline
PG 0007+106         &   2.6286&  10.9747& Q&  & 1.668&   1.815&$\pm$0.242&   2.195&$\pm$0.356&   4.235&$\pm$0.808&   9.39&$\pm$0.09\tabularnewline
NGC 34              &   2.7780& -12.1076& G&  & 0.398&   0.253&$\pm$0.017&   0.364&$\pm$0.024&   0.505&$\pm$0.122&  11.68&$\pm$0.05\tabularnewline
Arp 256N            &   4.7081& -10.3604& G&  & 0.549&   3.159&$\pm$0.016&   3.686&$\pm$0.016&   5.083&$\pm$0.037&   8.55&$\pm$0.00\tabularnewline
\end{longtable}
\tablefoot{\\
(1) Target name.\\
(2),(3) Far-infrared source position, as measured from the Gaussian fits.\\
(4) Class --  Galaxy (G) or QSO (Q).\\
(5) Morphology -- D denotes a double source with potentially inflated fitted size.\\
(6) Scale for the adopted distance and cosmology.\\
(7),(8),(9) Far-infrared half light radii at 70, 100, and 160\mum\/. Missing entries
for a given band indicate photometric S/N$<$10 or band not observed. Half
light radii are based on subtracting in quadrature observed width and PSF width.
The errors are statistical and do not include systematics due to the non-gaussianity
of the real source structure and PSF.\\
(10) Surface brightness of the 40-120\mum\ far-infrared emission.
}
\end{longtab}

\bibliographystyle{aa}
\bibliography{27706}

\end{document}